\documentclass[authoryear,12pt,5p,times]{elsarticle}
\usepackage{graphicx}
\usepackage{color}
\usepackage{amssymb,amsthm,amsmath}
\usepackage{aas_macros}
\usepackage{bm}

\def\apj{ApJ}

\usepackage{txfonts}
\usepackage[table]{xcolor}
\usepackage{booktabs}
\usepackage{tikz}
\usetikzlibrary{shapes,decorations,arrows,positioning}
\usepackage{verbatim}
\usepackage{lipsum}
\usepackage{hyperref}
\usepackage{mathtools}
\usepackage{listings}

\newcommand{\stargas}{$M_{\rm star}/M_\mathrm{gas}$}
\newcommand{\xmol}{x_\mathrm{mol}}
\newcommand{\fgas}{$f_\mathrm{gas}$}

\newcommand{\Zbin}{Z_\mathrm{bin}}
\newcommand{\R}{{\sf I\hspace*{-0.05cm}R}}
\newcolumntype{a}{>{\columncolor{gray!20}}c}
\newcolumntype{b}{>{\columncolor{white}}c}
\journal{Astronomy and Computing}
\begin{document}
%
%
\begin{frontmatter}
\title{The Overlooked Potential of Generalized Linear Models in Astronomy - I: Binomial Regression}
%
%
%

\author[MTA]{R S. de Souza}\ead{rafael.2706@gmail.com}
\author[Zoo]{E. Cameron}\ead{dr.ewan.cameron@gmail.com}
\author[USM]{M. Killedar}
\author[ASU,JPL]{J. Hilbe}
\author[HOU]{R. Vilalta}
\author[INAF,AIP]{U. Maio}
\author[SIS]{V. Biffi}
\author[MPA]{B. Ciardi}
\author[NOR]{J. D. Riggs}
\author{for the COIN collaboration}

\address[MTA]{MTA E\"otv\"os University, EIRSA ``Lendulet'' Astrophysics Research Group, Budapest 1117, Hungary}
\address[Zoo]{Department of Zoology, University of Oxford, Tinbergen Building, South Parks Road, Oxford, OX1 3PS, United Kingdom}
\address[USM]{Universit\"ats-Sternwarte M\"unchen, Scheinerstrasse 1, D-81679, M\"unchen, Germany}
\address[ASU]{Arizona State University, 873701,Tempe, AZ 85287-3701}
\address[JPL]{Jet Propulsion Laboratory, 4800 Oak Grove Dr., Pasadena, CA 91109}
\address[HOU]{Department of Computer Science, University of Houston
4800 Calhoun Rd,, Houston TX 77204-3010}
\address[INAF]{INAF --- Osservatorio Astronomico di Trieste, via G. Tiepolo 11, 34135 Trieste, Italy}
\address[AIP]{Leibniz Institute for Astrophysics, An der Sternwarte 16, 14482 Potsdam, Germany}
\address[SIS]{SISSA --- Scuola Internazionale Superiore di Studi Avanzati, Via Bonomea 265, 34136 Trieste, Italy}
\address[MPA]{Max-Planck-Institut f\"ur Astrophysik, Karl-Schwarzschild-Str. 1, D-85748 Garching, Germany}
\address[NOR]{Northwestern University, Evanston, IL, 60208, USA}


\begin{abstract}
 Revealing hidden patterns in astronomical data is often the path to fundamental scientific breakthroughs; meanwhile the complexity of scientific inquiry increases as more subtle relationships are sought. Contemporary data analysis problems often elude the capabilities of classical statistical techniques, suggesting the use of cutting edge  statistical methods. 
 In this light,  astronomers have overlooked a whole family of statistical techniques  for exploratory data analysis and robust regression, the so-called Generalized Linear Models (GLMs).
In this paper -- the first in a series aimed at illustrating the power of these methods in astronomical applications -- we elucidate the potential of a particular class of GLMs for handling binary/binomial data, the so-called  logit and probit regression techniques, from both a maximum likelihood and a Bayesian perspective.
As a case in point, we present the use of these GLMs to explore the conditions of star formation activity and metal enrichment in primordial minihaloes from cosmological hydro-simulations including detailed chemistry, gas physics, and stellar feedback. We predict that for  a dark  mini-halo with  metallicity  $\approx 1.3 \times 10^{-4} Z_{\bigodot}$, an increase of  $1.2 \times 10^{-2}$ in the gas molecular fraction, increases the probability of star formation  occurrence   by a factor of 75\%. Finally, we highlight   the use of   receiver operating characteristic  curves  as a   diagnostic for  binary classifiers, and ultimately we use these to demonstrate the competitive predictive performance of  GLMs against the  popular technique of artificial  neural networks.
\end{abstract}

\end{frontmatter}

%
\topmargin -1.3cm

\section{Introduction}
\label{sec:intro} 

The simple \textit{linear regression} model has long been a mainstay of astronomical data analysis, the archetypal  problem being to determine the line of best fit through Hubble's diagram \citep{hub29}.  In this approach, the expected value of the response variable, $\bm{Y} \in \R^m$, is supposed linearly dependent on its coefficients, $\bm{\beta} \in 
\R^n$, acting upon the set of  $n$ predictor variables, $\mathbf{X} \in \R^{n \times m}$, 
\begin{equation}
E(\bm{Y}) = (\bm{\beta}^T\mathbf{X})^T.
\label{eq:linear}
\end{equation}
The least-squares fitting procedure for performing this type of regression (\citealt{iso90}) relies on a number of distributional assumptions which fail to hold when the data to be modelled come from \textit{exponential family} distributions other than the Normal/Gaussian \citep{Hil12, hil14}. For instance, if the response variable takes the form of Poisson distributed count data (e.g.\ photon counts from a CCD), then the equidispersion property of the Poisson, which prescribes a local variance equal to its conditional mean, will directly violate the key linear regression assumption of \textit{homoscedasticity} (a common global variance independent of the linear predictors). Moreover, adopting a simple linear regression in this context means to ignore another defining feature of the Poisson: its ability to model data with only non-negative integers.
Similar concerns arise for modelling Bernoulli and binomial distributed data (i.e., on/off, yes/no) where regression methods optimized for continuous and unbounded response variables are of limited assistance \citep{Hil09}.

Yet, data analysis challenges of this sort  arise routinely in the course of astronomical research: for example, in efforts to characterize exoplanet multiplicity as a function of host multiplicity and orbital separation (Poisson distributed data; \citealt{wan14}), or to model the dependence of the galaxy bar fraction on total stellar mass and redshift (Bernoulli distributed data; \citealt{mel14}). For such regression problems there is a powerful solution already widely-used in medical research \citep[e.g.,][]{lindsey1999}, finance \citep[e.g.,][]{Jong2008}, and healthcare \citep[e.g.,][]{gri04} settings, but vastly under-utilized to-date in astronomy. This is known as Generalized Linear Models (GLMs). Basic GLMs include Normal or Gaussian regression, gamma and inverse Gaussian models, and the discrete response binomial, Poisson and negative binomial models.

\subsection{Generalized Linear Models}
The class of GLMs, first developed by \citet{nel72}, take a more general form than in Eq. \ref{eq:linear}:
\begin{equation}
E(\mathbf{Y}) = g^{-1}\left((\bm{\beta}^T \mathbf{X})^T\right),
\label{eq:glm}
\end{equation}
with the response variable, $\bf{ Y\ |\ \bm{\beta}^T X }$, belonging to a specified distribution from the single parameter exponential family and $g^{-1}(\cdot)$ providing an appropriate transformation from the linear predictor, $(\bm{\beta}^T \mathbf{X})^T$, to the conditional mean,  $\mu$.  The inverse of the \textit{mean function}, $g^{-1}(\cdot)$, is known as the {\it link function}, $g(\cdot)$.  \citet{nel72} and \citet{mcc89} laid the foundations of the GLM estimation algorithm, which is a subset of maximum likelihood estimation. The algorithm they devised in early software development is for the most part still used today in the majority of GLM implementations--both in commercial statistical packages (e.g.\ \textsc{SPSS} and \textsc{SAS}) and in freeware-type packages (e.g.\ \textsc{R} and \textsc{Python}).

GLMs have received a great deal of attention in the statistical literature.  Variations and extensions of the traditional algorithm have resulted in methodologies, such as:
generalized estimating equations \citep{lia86};
generalized additive models \citep{has86};
fixed and random effects regression \citep{bre93};
quasi-least squares regression \citep{sah14};
and more.
Bayesian statisticians working within the GLM framework have explored Gibbs sampling techniques for posterior sampling  \citep{alb93}, various issues of prior choice  \citep{gel08} and prior-sensitivity analysis \citep{dos94}, developed {\it errors-in-variables} treatments (for the case of errors in the predictor variables; e.g.\ \citealt{ric93,mal96}), and devised Gaussian process-based strategies for the use of GLMs in geospatial statistics \citep{dig02}.
The GLM methodology thus stands at the base of a wide number of contemporary statistical methods.

Despite the ubiquitous nature of GLMs in general statistical applications, there have been only a handful of astronomical studies applying  GLM techniques such as logistic regression (e.g.\ \citealt{Rai12,Rai14,lan14}),  Poisson regression (e.g.\ \citealt{and10}); and the importance of modelling overdispersion in count data (as facilitated by the negative binomial GLM) has only lately become appreciated through cosmological research \citep{ata14}.  Hence, in this series of papers we aim to demonstrate the vast potential of GLMs to assist with both exploratory and advanced astronomical data analyses through the  application to a variety of astronomical inference problems.

The astronomical case studies explored herein focus on an investigation of the statistical properties of baryons inside simulated high-redshift haloes, including detailed chemistry, gas physics and stellar feedback.
 The response variables are categorical with two possible outcomes and therefore Bernoulli distributed. In our particular case, these correspond to either (i) the presence/absence of star formation activity, or (ii) metallicity above/below the critical metallicity ($Z_{\rm crit}$) associated with the first generation of stars. The predictor variables are properties of high-redshift galaxies with continuous domain.

The outline of this paper is as follows. In \S\ref{sec:data} we describe the cosmological simulation and the dataset of halo properties. We describe various forms of binomial GLM regression in \S\ref{sec:GLM}. In \S\ref{sec:application} we present our analysis of the simulated dataset for the two selected response variables. In \S\ref{sec:diag} we discuss critical diagnostics of our analysis, and compare our classifications with those that use artificial neural networks in \S\ref{sec:comp}. Finally,  in \S\ref{sec:end} we summarize our conclusions.

\section{Simulations}
\label{sec:data}
In order to ascertain the key ingredients that affect star formation in the early Universe, we study cosmological simulations of high-redshift galaxies and proto-galaxies. In the following, we describe the simulated data used to exemplify the unique benefits of binomial GLM regression for modelling galaxy properties that are naturally addressed as a dichotomous problem.

\subsection{Runs}
\label{subsec:runs}

The data set used in this work is retrieved from a cosmological 
hydro-simulation based on \citealt{Biffi2013} (see also \citealt{maio2010, maio2011a, desouza2014}).
The code employed to run the simulation is \textsc{gadget-3}, a modified version of the parallel $N$-body, smoothed-particle hydrodynamics code named \textsc{gadget-2} \citep{Springel2005}.
The modifications include: a relevant chemical network to self-consistently follow the evolution of different atomic and molecular chemical species 
\cite[e.g.,][]{yoshida2003, maio2006, maio2007, maio2009};
metal pollution according to proper stellar yields
and lifetimes for both the pristine population III (Pop III) and the following population  II/I (Pop II/I) star forming regime \cite[][]{tornatore2007, maio2010}; radiative gas cooling from molecular, resonant and fine-structure lines \cite[][]{maio2007}.
The actual stellar population is determined by the local heavy-element mass fraction (metallicity, $Z$) and the existence of a critical threshold $Z_{\rm crit} = 10^{-4}Z_{\bigodot}$\footnote{Despite the uncertainties on $Z_{\rm crit}$, it is safe to assume values around $Z_{\rm crit} = 10^{-4}Z_{\bigodot}$, in fact even order-of-magnitude deviations would not change significantly the final results in terms of star formation and cosmic metal pollution \cite[see details in][]{maio2010}.} \citep[e.g., ][]{Omukai2000, Bromm2001} below which Pop III star formation takes place and above which Pop II/I stars are formed.

The initial matter density field is sampled at redshift $z = 100$ adopting the standard cold dark matter model with cosmological constant $\Lambda$, $\Lambda$CDM.
The cosmological parameters at the present time are assumed to be: $\Omega_{0,\Lambda} = 0.7$, $\Omega_{0,m} = 0.3$, $\Omega_{0,b} = 0.04$, for cosmological-constant, matter and baryon density, respectively \citep[e.g.,][]{Komatsu2011}. 
The expansion parameter at the present day is assumed to be $H_{0}= 100 \,h\,\rm km  s^{-1} Mpc^{-1}$, with $h=0.7$, while the primordial power spectrum has a slope $n=1$ and is normalized by imposing a mass variance within the 8-kpc/$h$ sphere radius of $\sigma_8 = 0.9$.
We consider snapshots in the range $9 \lesssim z \lesssim 19$, for a cubic volume of comoving side $\sim$0.7 Mpc, sampled with $2 \times 320^{3} $ particles per gas and dark-matter species. The resulting resolution is $\rm 42~M_{\bigodot} {\it h^{-1}}$ and $\rm 275~M_{\odot} {\it h^{-1}}$ for gas and dark matter, respectively.

\subsection{Data set}\label{subsec:data}
The simulation outputs  considered here  consist of six parameters: dark-matter mass, $M_{\mathrm{dm}}$,
gas mass, $M_{\mathrm{gas}}$,
stellar mass, $M_{\mathrm{star}}$,
star formation rate, \textit{SFR}, 
metallicity, $Z$,  
and gas molecular fraction, $\xmol$. In addition to the data-set described above, we incorporate in the analysis the following derived quantities:
gas fraction, 
\fgas$\equiv M_{\mathrm{gas}}/M_{\mathrm{dm}}$,
stellar fraction, 
$f_\mathrm{star} \equiv  M_{\mathrm{star}}/M_{\mathrm{dm}}$
and stellar-to-gas mass ratio \stargas.

The  sample studied in this work is  composed of 1680 haloes in the whole  redshift range, with about $200$ objects at $z = 9$.
The masses of the haloes are in the range
$10^{5} M_{\bigodot} \lesssim M_{\rm dm} \lesssim 10^{8} M_{\bigodot}$, with corresponding gas masses between $10^4 - 10^7 M_{\bigodot}$. Table \ref{tab:dataset} summarizes the statistics  of the halo parameters contained in the sample. 
The interested reader can find in \cite{Biffi2013} a more detailed discussion of the thermal and dynamical properties of the primordial objects analysed in this paper.

\begin{table*}

\begin{center}
\begin{tabular}{l  c   c  c  c  }
\hline
\rowcolor{cyan!20}
Variable name & Minimum  & Maximum  &  Mean &  Standard deviation  \\
\hline
\rowcolor{gray!10}
Dark-matter mass: $M_{\mathrm{dm}}~(M_{\bigodot})$ & 2.20 $\times 10^{5}$ & $5.59 \times 10^{7}$& $2.15 \times 10^{6}$ & $4.39 \times 10^{6}$\\
Gas mass: $M_{\mathrm{gas}}$ ($M_{\bigodot}$) & $1.27 \times 10^{4}$ & $5.80 \times 10^{6}$  & $1.39 \times 10^{5}$ & $4.18 \times 10^{5}$  \\
\rowcolor{gray!10}
Stellar mass: $M_{\mathrm{star}}~(M_{\bigodot})$  & 0  & $3.45 \times 10^{4}$ & $2.87 \times 10^2$ & $2.42 \times 10^3$  \\
Star formation rate: \textit{SFR} $(M_{\bigodot}/yr)$ & 0 & $3.08\times 10^{6}$ & 2.17 $\times 10^{4}$ & $1.70 \times 10^{5}$  \\
\rowcolor{gray!10}
Metallicity: $Z$ ($Z_{\bigodot}$) & 0 & $1.03 \times 10^{-2}$ & $1.28 \times 10^{-4}$& $8.49 \times 10^{-4}$ \\
Gas molecular fraction: $\xmol$  &7.53 $\times 10^{-6}$ & $1.31 \times 10^{-1}$  & $2.20 \times 10^{-3}$ & $1.18 \times 10^{-2}$ \\
\rowcolor{gray!10}
Gas fraction: \fgas$\equiv M_{\mathrm{gas}}/M_{\mathrm{dm}}$  & $1.66 \times 10^{-2}$   & $1.21 \times 10^{-1}$ & $4.22 \times 10^{-2}$ & $1.87 \times 10^{-2}$ \\
Stellar  fraction: $f_\mathrm{star} \equiv  M_{\mathrm{star}}/M_{\mathrm{dm}}$  & 0 & $1.06 \times 10^{-3}$  & $1.39\times 10^{-5}$  & $7.92\times 10^{-5}$\\
\rowcolor{gray!10}
Stellar-to-gas mass ratio: \stargas  & 0 & $1.70 \times 10^{-2}$    & $1.86 \times 10^{-4}$  & $1.14 \times 10^{-3}$ \\
\hline
\end{tabular}
\caption{Summary statistics  of the halo properties. }
\label{tab:dataset}
\end{center}

\end{table*}

\section{GLM Regression for Binary Response Data}\label{sec:GLM}
%
In preparation for the application  of binomial GLM regression  we begin with a discussion of the two most common link functions: logit and probit (\S \ref {logit}). Then we describe   three variations on a class of GLMs which apply to binary response data: the maximum likelihood estimation (MLE) approach with logit link function (\S \ref{maxlikeGLM}); and the Bayesian approach with a logit link function (\S \ref{bayesLR})  and with a probit link function amenable to exact Gibbs sampling (\S \ref{bayesPR}). These will be applied in the following section (\S\ref{sec:application}) in the context of two specifically chosen astrophysical problems: i) presence/absence of star formation activity; ii) gas metallicity below/above $Z_{\rm crit}$ to discriminate between Pop III/Pop II/I star formation mode. 
The interested reader can find a comprehensive description of the underlying theory behind  GLMs  in \citet{zuu13}.

\subsection{Logit and probit regression}
\label{logit}
\
The Bernoulli distribution describes a process  in which there are only two possible outcomes: success or failure (yes/no, on/off, red/blue, etc.; typically coded as 1/0)--the former occurring with probability, $p$, and the other with probability, $1-p$.  For multiple independent Bernoulli observations the total \textit{success} count, $k$, follows a binomial distribution\footnote{See\ \citet{cam10} for a review of the binomial distribution and both maximum likelihood  and Bayesian approaches to estimation of confidence/credible intervals on $p$.}, $P(k) =  \binom {n} {k} p^k (1-p)^{n-k}$.  Both distributions are members of the exponential family (supposing the number of binomial trials, $n$, is known and fixed) and thus may be used (equivalently) as the response distribution for modelling binary response data in the GLM framework.

\begin{figure}[t]
\centering
\includegraphics[width=1\columnwidth]{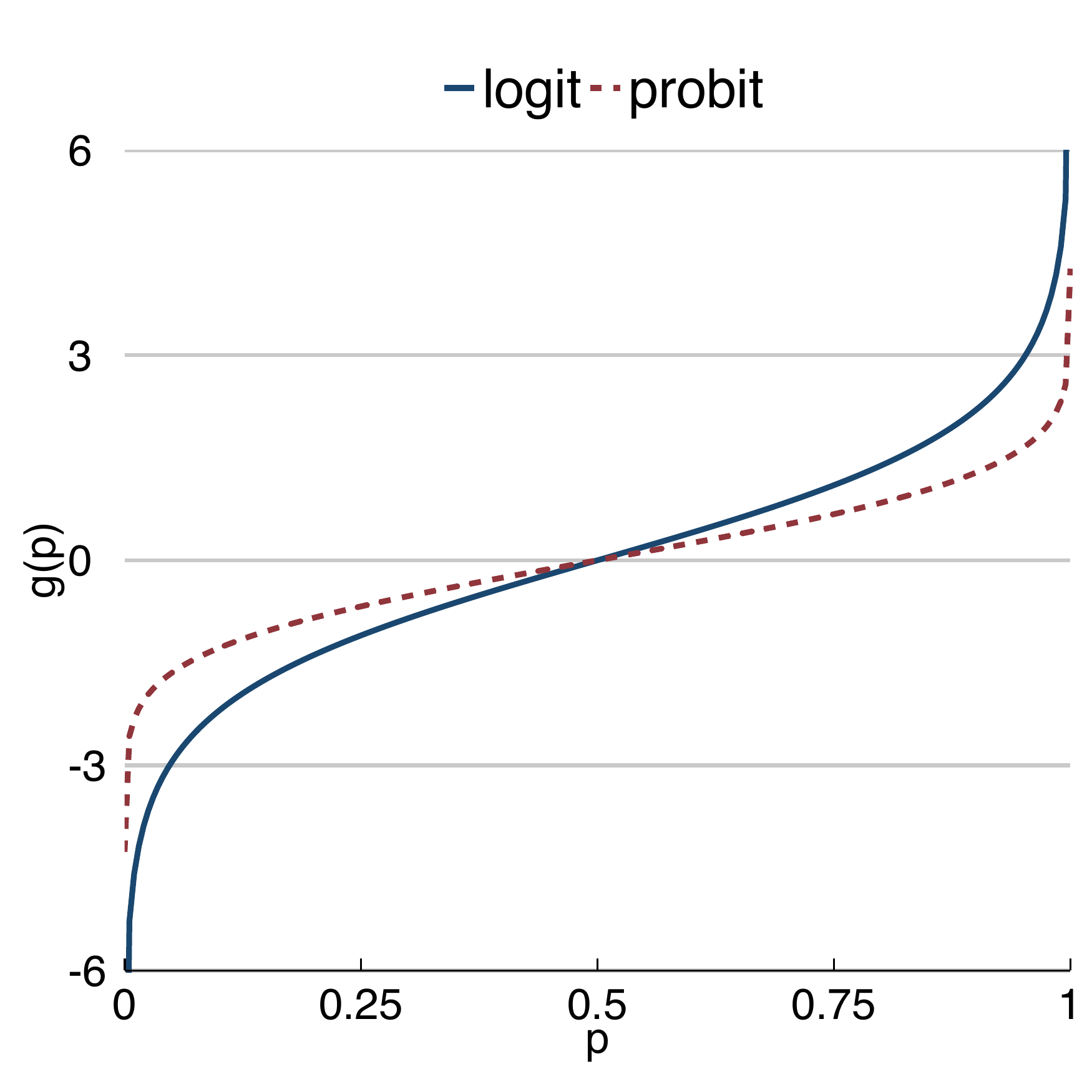}
\caption{Comparison between logit, solid blue curve and probit, dotted red curve,  link functions g(p).}
\label{fig:links}
\end{figure}

The link function chosen in this case is designed to ensure a bijection\footnote{A function $f$ from a set $X$ to a set $Y$ with the property that, for every $y$ in $Y$, there is exactly one $x$ in $X$ such that $f(x) = y$.} between the $(-\infty,\infty)$ range of 
the linear predictor, $(\bm{\beta}^T \mathbf{X})^T$,  and the (0,1) range of non-trivial probabilities for the binomial population proportion (the Bernoulli $p$).
To this end there are two popular choices:
the {\it logit} function, 
\begin{equation}
\label{eq:logit}
g(p) = \log \frac{p}{1-p},
\end{equation}
and the {\it probit} function, 
\begin{equation}
\label{eq:probit}
g(p) = \Phi^{-1}(p), 
\end{equation}
where $\Phi(\cdot)$ represents the Normal distribution function.  The choice of  link function defining the \textit{logit} predicted value, $\mu$,
\begin{equation}
\bm{\mu}^T = g^{-1}(\bm{\beta}^T \mathbf{X}) = \frac{\exp( \bm{\beta}^T \mathbf{X})}{1 + \exp(\bm{\beta}^T \mathbf{X})},
\end{equation}
or the \textit{probit} predicted value,
\begin{equation}
\bm{\mu}^T = g^{-1}(\bm{\beta}^T \mathbf{X}) = \Phi(\bm{\beta}^T \mathbf{X}), 
\end{equation}
accordingly.  Both link functions describe sigmoid curves smoothly and monotonically increasing from $\mu=0$ at $\bm{\beta}^T \mathbf{X} = -\infty$ to $\mu=1$ at $\bm{\beta}^T \mathbf{X} = \infty$ with the greatest rate of change occurring at $\bm{\beta}^T \mathbf{X} = 0$,  as displayed in Fig \ref{fig:links}.

The logit function is most commonly preferred in clinical research applications where outcomes are most naturally described in terms of the odds-ratio, $\frac{p}{1-p}$ (e.g.\ the relationship between the odds-ratio of patient recovery/non-recovery and the concentration of an administered drug); whereas the probit function is often presented within Bayesian statistical applications exploiting an associated Gibbs sampling algorithm. Sigmoid curves such as those described by the logit and probit functions may already be seen in empirical/phenomenological astronomical models: for example, in describing the fraction of quenched galaxies as a function of mass and/or environmental density \citep{pen10,rod14}.

A reason for employing logit or probit regression to model binary response data is to obtain for objects with only $\mathbf{X}$ observations, but no observed $\mathbf{Y}$'s, the predicted probabilities that the unobserved response variable has the value 1 indicating ``success'', however that is defined (e.g., "galaxy is quenched", "star hosts planet").  Both models usually produce similar probabilities; though probit regression is not as commonly used for assessing the relationship of a predictor to the response since the interpretation of the exponentiated coefficient of a logit predictor as an odds ratio is a desirable feature of that model. Probit regression is normally used when a continuous variable is dichotomized so that it becomes a binary response \citep[see][for an examination of these and related issues with logit and probit models from both a frequency and Bayesian perspective] {zuu13}. It is worth noting that,  while other well known machine learning algorithms (e.g., support vector machines, k-nearest neighbourhood) can be used for  binary classification tasks, their main focus is prediction rather than modelling.  In other words,  their aim  is to find a functional  algorithm $f(x)$ that operates on $x$ to  predict the responses $y$. GLMs, on the other hand,  represent a data modelling  philosophy, which assumes  the sample  to be generated by a stochastic process, e.g., Gamma, Poisson, with   Bernoulli being the case  study here. While the former may lead to a more accurate prediction for  complex problems, although there are plenty of GLM extensions for such cases,   the later allows a clearer  interpretation of the relationship between the predictor  and response variables (see \S \ref{sec:application}).

\subsection{Maximum-likelihood estimation GLM regression with logit link function}
\label{maxlikeGLM}
Despite the growing popularity of Bayesian statistical analysis  in the physical  sciences the MLE approach to GLM fitting remains the default in the majority of statistical software packages\footnote{Such as the \texttt{glm}  in {\sc R}}: for this reason, and its historical significance (cf.\ the extensive treatment given by \citealt{mcc89}), we describe this approach first.

With the likelihood of the  dataset  fully specified by the linear predictor, $\bm{\beta}^T\mathbf{X}$, and the choice of response variable distribution and link function of the GLM, the corresponding likelihood function for regression is both readily tractable and easily evaluated computationally.  Iterative algorithms operating on the negative log-likelihood, such as the iteratively re-weighted least squares  procedure used by \texttt{glm} \citep{ven02}, thus provide a fast computational strategy for recovering the MLE solution.  The output from a standard MLE GLM fitting code will typically be a list containing: (i) a MLE estimate, $\hat{{\beta}}_i$, for the ${\beta}_i$ component of each candidate predictor variable, $\mathbf{X}_i$; (ii) the associated estimate of its standard error, $\hat{\sigma}_{{\beta}_i}$, from which approximate confidence intervals (CI) on ${\beta}_i$ may be obtained using the Normal distribution function (e.g.\ a 95\% CI: $\hat{{\beta}}_i \pm 2 \hat{\sigma}_{{\beta}_i}$); and (iii) a $p$-value computed from the Wald test using (i) and (ii), required for significance testing of the given predictor variable. The Wald test determines how significant a predictor  variable is,  where for the   GLM case  it tests  the predictor parameter values, $\hat \beta$, versus 
hypothesized values, $\beta_0$ (often $0$ for logistic models), and is based on MLE. The difference between $\hat \beta$ and $\beta_0$ divided by the standard error, $se$,  of the residuals follows an approximate Gaussian distribution. For a specific estimate, we have 
\begin{equation}
\frac{\hat \beta - \beta_0}{se(\hat \beta)} \overset{.}{\sim} \mathcal{N} (0,1) ,
\end{equation}
where $se(\hat \beta)$ is the standard error of the estimate of $\beta$. A Wald 95\% confidence interval for $\hat \beta$ is given by 
\begin{equation}
\hat \beta \pm 1.96 se(\hat \beta), 
\end{equation}
\citep[see e.g.,][]{Pawitan2001,Hil09}.
Estimation of (ii) is by way of the observed information matrix according to asymptotic convergence theory for MLE estimation.

In {\sc R} the {\sc glm} procedure may be called to perform MLE estimation of the logistic regression model using the general syntax shown  in  \ref{sec:scripts}.

\subsection{Bayesian GLM regression with logit link function}
\label{bayesLR}

The {\sc bayesglm} function in the CRAN\footnote{\href{http://cran.r-project.org}{http://cran.r-project.org}} {\sc arm} package is commonly used to estimate Bayesian logistic models
\citep{arm}.
 The code used to estimate this class of models is based on {\sc R's} default {\sc glm} function (see  \ref{sec:scripts}).
Normal and Jeffreys priors have traditionally been favoured for use with continuous predictors in logit regression (e.g.\ \citealt{raf96,ibr91}); though more recently the Cauchy has been strongly promoted as an optimal default prior \citep{gel08}.  It is also recommended that  continuous  predictors be centered if not fully standardized\footnote{E.g., transformed to zero sample mean and unit sample variance as $x^* = (x-\hat\mu_x)/\hat \sigma_x$, where $x^*$ is the standardized variable, and $\hat \mu_x$ and $\hat \sigma_x$ represent the sample mean and standard deviation respectively.}, if the predictor is not linearly related to the response; withal,  standardizing continuous predictors help convergence, when entered into a Bayesian GLM,  since it puts them on the same scale.  Care must always be taken to assure that a default prior's use with the data makes sense; to this end visual inspection of mock datasets generated from the prior--likelihood pairing unconstrained by the data can serve as an effective  check.  To this end ``functional uniform'' priors provide another means to limit prior-sensitivity in the shape of the preferred fitting function; cf.\ \citealt{bor12}.  For the purposes of the present study we follow the Cauchy prior recommendations of \citet{gel08}.

\subsection{Bayesian GLM regression with probit link function}
\label{bayesPR}
\
Use of the probit link function for Bayesian GLM regression has become a popular choice owing to the availability of an exact Gibbs sampling algorithm for this model presented by \citet{alb93}.  The novelty of their algorithm is a data augmentation scheme in which an additional latent variable is added for each observation having standard Normal distribution with mean set by the linear predictor, from which the likelihood of the observed response is determined according to whether or not this latent variable is above or below zero.  Although the general sampler of the \texttt{arm} package does not in fact implement the \citet{alb93} scheme, it is important to note its availability for use in more complex Bayesian hierarchical models built on the GLM framework (e.g.\ for the case of errors-in-variable GLM regression with binary distributions for both predictor and response variable, such as can arise in comparing the sensitivity of two alternative tests)\footnote{An alternative \textsc{R} function implementing the data augmentation scheme for Bayesian probit regression is available in the CRAN \textsc{LearnBayes} package \citep{alb07} as \texttt{bayes.probit}.}. 
The basic syntax for using Bayesian probit GLM in \textsc{R} is summarized in \ref{sec:scripts}. 
The same criteria for the use of priors that we discussed for logit models above also maintain for probit models. However, if the analyst desires to interpret the coefficients in terms of odds or risk ratios, a logit model must be used, regardless if the model is based on MLE or Bayesian methods.

\section{Application to cosmological simulations}
\label{sec:application}
Within this section we demonstrate the application of the binomial regression techniques introduced above to answer questions from an exploratory analysis of our cosmological hydro-simulation dataset that could not be addressed by standard linear regression methods. This is because probability of occurrence  for a binary outcome is bounded between 0 and 1,  while  the underlying theory of linear regression allows realizations with values out of this range. 
Rather than exhaust all possible techniques for a single dataset, our aim is to demonstrate practical differences between distinct
 types of binary regression: i) Bayesian vs MLE approach, both with the standard logit link (\S \ref{maxlikeGLM},\ref{bayesLR});  ii) Bayesian regression comparing logit vs probit link functions (\S \ref{bayesLR},\ref{bayesPR}). In the first case we consider the star formation activity connection with a preselected (physically motivated) set of predictor variables: $\xmol$ and $Z$. Alternatively, in our later analysis of the metallicity content of the galaxies, we use an automatic criterion to select the best choice of
 predictor variables among the entire set of halo features, or in other words the variable combination that minimizes the Akaike Information Criterion \citep[AIC;][]{Akaike1974}. For all the following GLM analyses we quote the maximum likelihood (${\cal L}_\mathrm{max}$),  the AIC as well as the alternative Bayesian Information Criterion \citep[BIC;][]{schwarz1978}.

\subsection{Star formation activity }\label{subsec:P1}

Here we discuss the connection between star formation activity and the gaseous chemical properties, $\xmol$ and $Z$ of proto-galaxies, using a Bayesian and a MLE approach with logit link.
The formation of the first metal-free stars in the Universe ended the cosmic dark ages \citep[][]{desouza2011, desouza2012, Bromm2013, desouza2013, Whalen2013, Whalen2013b} and began the production of elements heavier than lithium \cite[][]{maio2010, maiodla2013, wise2014}. Thus, a key problem in physical cosmology is to understand the environmental properties  of such objects \citep[e.g.,][]{desouza2013b, Biffi2013, salvaterra2013}, born out of the pristine conditions leftover by the Big Bang.

\begin{figure}[t]
\centering
\includegraphics[width=1\columnwidth]{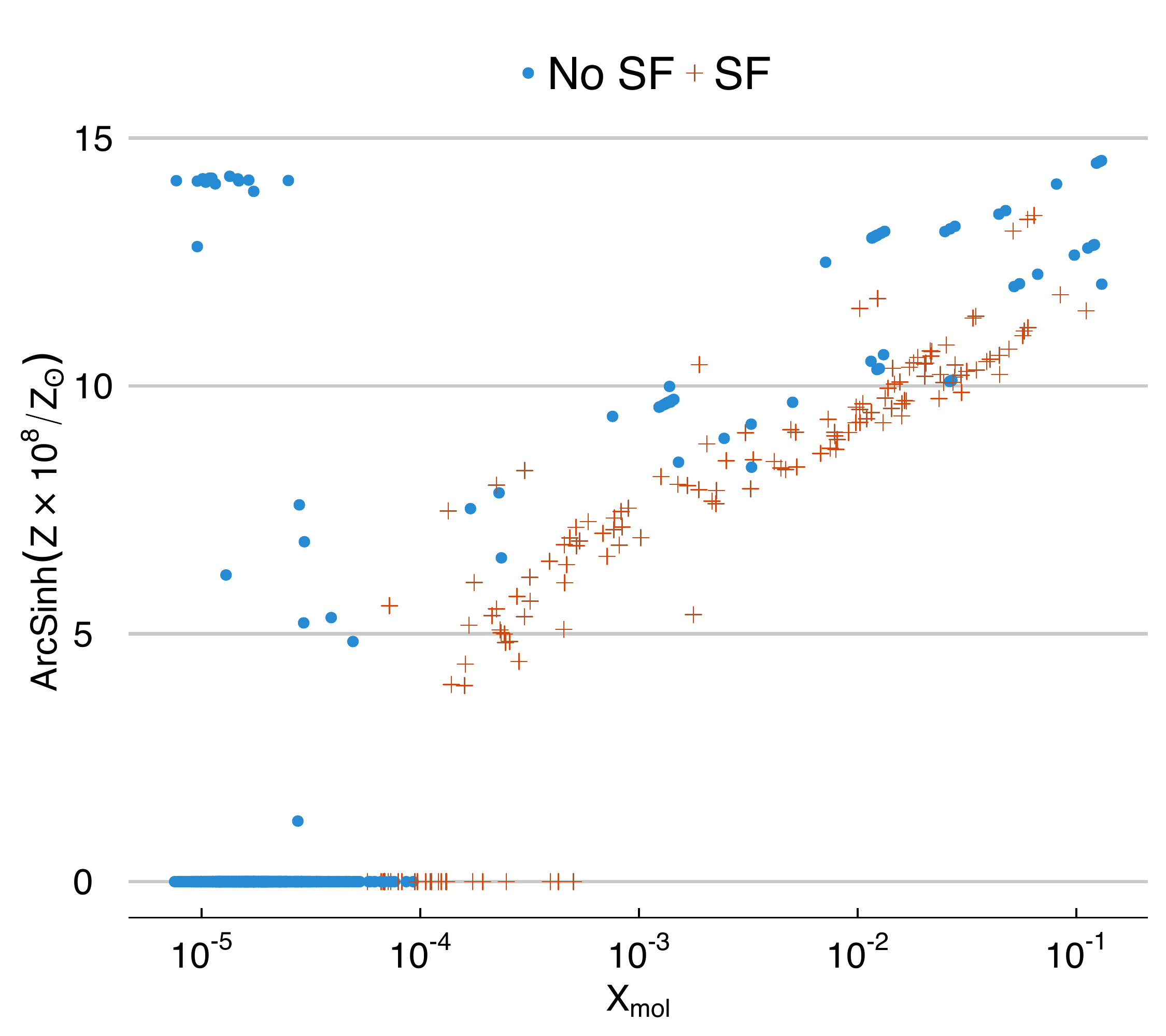}
\caption{Molecular fraction and gas metallicity  for all  haloes in the simulation, colour-coded by presence of star-formation  activity: blue dots indicate  no SF and red crosses indicate SF. 
The re-scaling and  ArcSinh transformation to $Z$ is  done in order to allow a better visualization  of the whole   range of metallicities,   including the null values.}
\label{fig:Scat-P1}
\end{figure}

As a visual exploration, Figure \ref{fig:Scat-P1} shows the scatter of $\xmol$ and $Z$ coloured according to the presence of star formation activity.
The objects located in the top left with  high $Z$ and very low $x_{\rm mol}$ are  strongly displaced from the general trend, highlighting the effects of metal enrichment of quiescent galaxies polluted by external sources.
The bottom right corner is not populated because gas with large molecular fractions of $\xmol \sim 10^{-2}$ or higher would have very short cooling times, hence would immediately form stars which pollute the surrounding medium. Therefore, the larger the deviation from the general trend, the higher the effects of feedback mechanisms.

\begin{figure}[t]
\centering
\includegraphics[width=1\columnwidth]{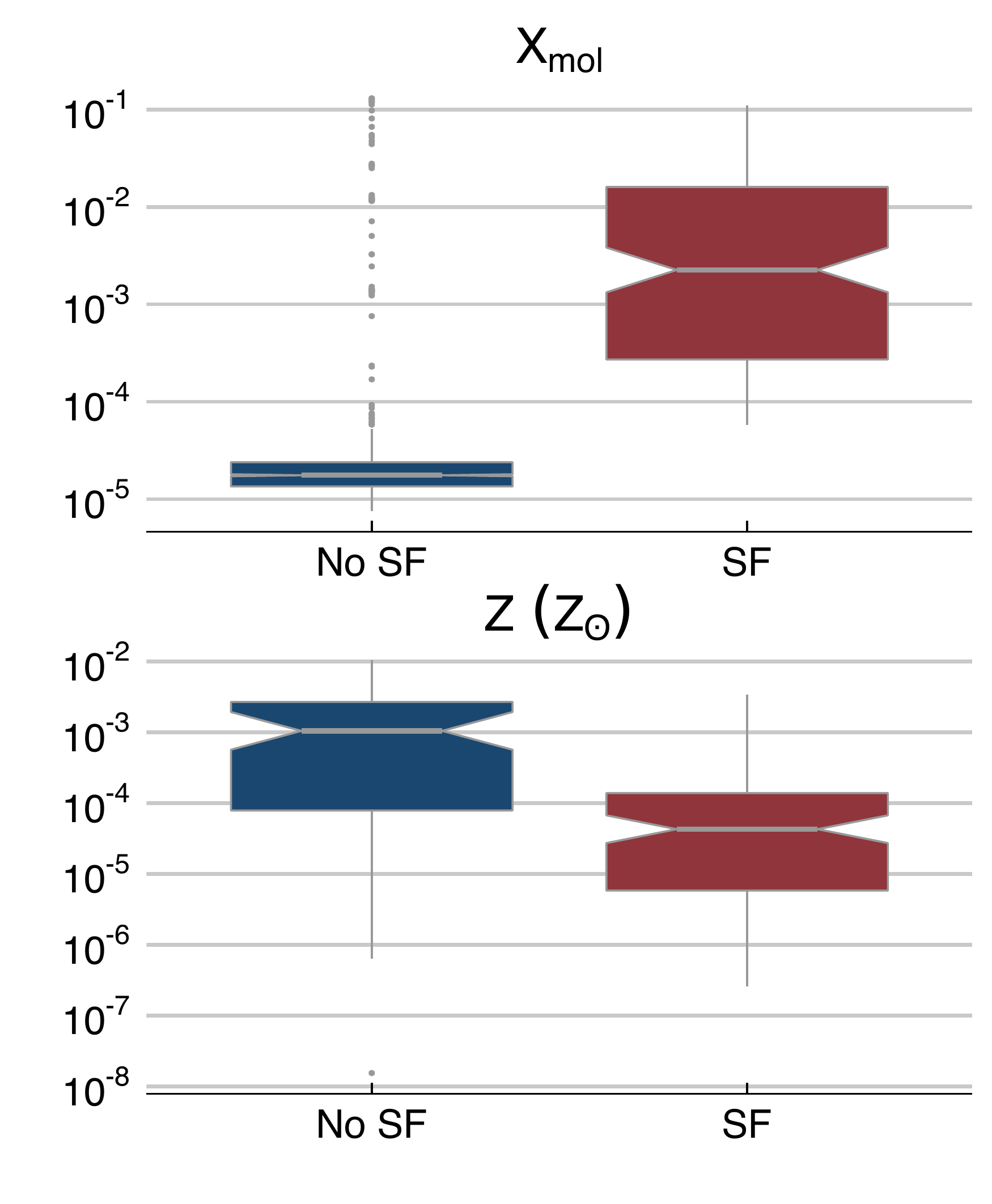}
\caption{Distribution of molecular fraction and metallicities for haloes 
 colour-coded by whether they host star-formation activity or not. Red colour represents haloes with SF and blue colour haloes with no  SF. The bottom and top of the box  show  the first and third data quartiles, while the band inside the box their median. The  notches represent  a rough guide of the uncertainty around the median of each distribution, $\pm 1.58\times \rm{IQR}/\sqrt{n_{obj}}$, with $n_{obj}$ being the number of objects, and IQR standing  for interquartile range. }
\label{fig:notch}
\end{figure}

Figure  \ref{fig:notch} represents  the distribution of $\xmol$ and $Z$  colour-coded by star formation activity and displayed by a box plot. The  notches represent  a rough guide of the uncertainty around the median of each distribution, $\pm 1.58\times \rm{IQR}/\sqrt{n_{obj}}$, with $n_{obj}$ being the number of objects, and IQR standing  for interquartile range. 
A visual inspection suggests  that $\xmol$ plays a major  role in triggering the star formation activity, in contrast to the lower influence of $Z$ 
\citep[see e.g., ][]{desouza2014}. 
The medians of haloes 
with and without star formation are different for both $\xmol$ and $Z$, indicating  they might represent different populations, which reinforce their choice as predictor variables for star formation activity.

\begin{table}[t] \centering
	
	\caption{$\hat{\beta}_i$ coefficients from  MLE and Bayesian (with Cauchy prior) GLM logit regression analysis with $\textit{SFR}_{\rm bin}$ as the response variable and $\xmol$ and $Z$ as  predictors. The associated $p$-values ($\cal P$, see\S \ref{maxlikeGLM}),  are listed underneath the coefficients. The logarithm of the maximum likelihood (${\cal L}_\mathrm{max}$), the AIC and the BIC for each choice of link function are also shown. }
 \label{tab:MLBayeslogit}
   
\begin{tabular}{llcc} 
\\[-2.2ex]\hline 
\hline \\[-2.2ex] 
 			& & \multicolumn{2}{c}{\textit{Response variable:}} \\ 
\cline{3-4} 
\\[-2.2ex] 	& & \multicolumn{2}{c}{$\textit{SFR}_{\rm bin}$} \\ 
\\[-2.2ex] 	& & MLE logit & Bayes logit\\ 
\hline \\[-2.2ex] 
\rowcolor{gray!10}
$\hat{\beta}_{0}$ & & $- 2.51 \pm 0.10$ & $-2.50 \pm 0.10$ \\ 
\rowcolor{gray!10}
& & ($\cal P \ll$ 0.001) & ($\cal P \ll$ 0.001) \\ 
  & & \\ [-2.2ex]
$\hat{\beta}_{1}$ & $Z$ & $-1.15 \pm 0.33$ & $-1.05 \pm 0.27$ \\ 
  & & $(\cal P$ = 0.0006) & $(\cal P$ = 0.0001) \\ 
  & & \\ [-2.2ex]
\rowcolor{gray!10}  
$\hat{\beta}_{2}$ & $\xmol$ & $1.05 \pm 0.15$ & $1.01 \pm 0.14$ \\ 
\rowcolor{gray!10}
& & ($\cal P \ll$ 0.001) & ($\cal P\ll$ 0.001) \\ 
  & & \\ [-2.2ex] 
\hline \\[-2.2ex] 
${\cal L}_\mathrm{max}$%
				& & $-452$ & $-452$ \\ 
AIC				& & 909 & 909 \\ 
BIC				& & 926 & 926 \\ 

\hline 
\hline \\[-2.2ex] 
\end{tabular} 
\end{table}

To perform the GLM analysis, we  categorize the haloes via the binary response variable $\textit{SFR}_\mathrm{bin}$, as those with  ({\it SFR} $>0$ ) and without  ({\it SFR} = 0) star formation activity, a binary classification which makes it suitable for a binomial GLM analysis,
\begin{equation}
\textit{SFR$_\mathrm{bin}$} = \left\{
 \begin{array}{rlr}
  1 & \text{or `SF'~} 		&\text{if }\textit{SFR} > 0,\\
  0 & \text{or `no SF'~}	&\text{if }\textit{SFR} = 0.
 \end{array} \right.
\end{equation}
The underlying properties that act as predictor variables are: $\xmol$ and $Z$. 
We indicate with $p$ the  probability that star formation activity is occurring in a galaxy. More specifically, $p=1$ (0) if a galaxy has (has no) star formation. The predicted probability $\pi$ is then determined by the GLM analysis and compared  to observed probability $p$ (for a given decision boundary), in order to ascertain the method's performance,  as explained below. We standardized the predictors  before the GLM analysis in order to ameliorate possible collinearity  and  scaling bias due to units differences.

Table~\ref{tab:MLBayeslogit} shows  the estimated coefficients  and   related $p$-values\footnote{$p-$values measure the significance of the term  associated to the fitted coefficients, $\hat{\beta_i}$. $p$-values $\leq$ 0.05 imply that $\hat{\beta_i}$ are significant at  least at the 95\% confidence level. To avoid possible confusion with the probability $p$ we indicate the $p$-values as $\cal P$.} for the various linear predictors for both Bayesian and MLE approach with the standard logit link. 

\begin{figure}[t]
\centering
\includegraphics[trim=1.5cm 2cm 2cm 2cm, clip=true, width=1\columnwidth]{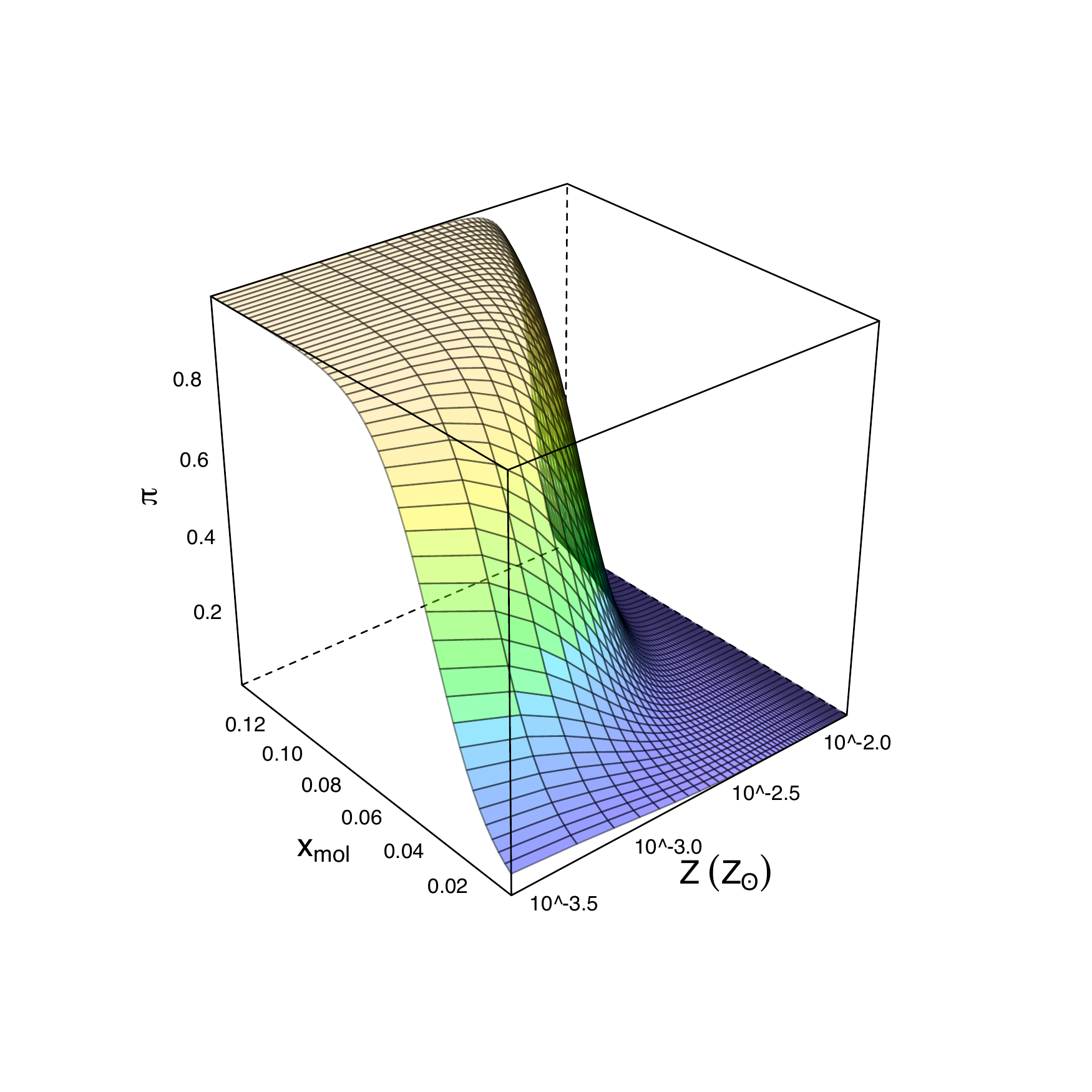}
\caption{Predicted probabilities, $\pi$,  of star formation  activity  vs metallicity, $Z$,  and molecular fraction, $x_{\rm mol}$,  for the  logit regression. }
\label{fig:logit3D}
\end{figure}

The coefficients for the logit model represent the log of the odds ratio for \textit{SF} activity. Since  the predictors are scaled, it allows for performing a relative   comparison between   variables measured in different units. A one $\sigma$ increase in the  halo  metallicity ($\approx 8.5 \times 10^{-4} Z_{\bigodot}$) produces,  on average, a change of $-1.15$ in the log of odds ratio ($\approx 25\%$  in probability)  for presence of \textit{SF}, for an average halo with gas molecular fraction close to the mean,   $x_{\rm mol} \approx 2.2\times 10^{-3}$.  Likewise, for a halo with   $Z \approx 1.3 \times 10^{-4} Z_{\bigodot}$, an increase of  $1.2 \times 10^{-2}$ in   $x_{\rm mol}$, increases the probability of \textit{SF}   by  75\%. The analysis  not  only confirms     $x_{\rm mol}$  as a critical   parameter to trigger  the   \textit{SF} in primordial halos, in agreement with previous  works \citep[see e.g,][]{desouza2014},   but provides the means to interpret  the role of each halo property  in terms of odds and  probabilities.
As stated in \S \ref{sec:GLM}, the GML analysis provides an estimate $\hat{\beta}_i$ for the $\beta_i$ component of each predictor variable. The values obtained can be used to calculate the linear predictor, $\eta$:
\begin{equation}
\eta = \hat{\beta}_0+\hat{\beta}_1 Z +\hat{\beta}_2 \xmol,
\end{equation}
and transformed into a predicted  probability, $\pi$:
\begin{equation}
\label{eq:pi}
\pi = \frac{e^{\eta}}{1+e^{\eta}},
\end{equation}
which uses the logit link defined in Eq. \ref{eq:logit}. Figure  \ref{fig:logit3D} displays the regression plane solution using the logit link. The surface gives the probability  of \textit{SF} activity for each pair ($Z$, $x_{\rm mol}$).

This can be used to assign a class membership for each object for a given probability decision threshold, $\pi_{th}$,   i.e. SF = 1 if $\pi > \pi_{th}$  and 0 if $\pi < \pi_{th}$. 
For each halo, the predicted probability can be  compared to the observed probability,  which in this case is $p$ = 1 if the halo presents star formation activity, and $p$ = 0 otherwise.  The performance of the method in reproducing the correct observed probabilities can be evaluated as detailed in \S \ref{sec:diag}.
When class sample sizes are approximately equal, which in this scenario would  imply  a similar number of galaxies with and without star formation activity,  the optimal decision threshold is $\pi_{th} \sim 0.50$ (see \S \ref{subsec:ROC}).

Nevertheless, this criterion  is  not appropriate when the class sizes are imbalanced  and an adjusting decision threshold has to be used. As a trivial example, if the data is imbalanced,  the fit can   predict  $\pi = 0.2$ for all haloes with \textit{SF} = 1 and  $\pi = 0.1$ for all haloes with SF = 0. In this hypothetical  scenario, the decision boundary would be in the range 0.1-0.2, instead of being 0.5 (50\%) as one would naively expect.  
A more detailed explanation of how to adjust the decision threshold probability,  $\pi_{th}$, and a discussion of the predictive power of the method  is given in \S \ref{sec:diag}.  

The MLE and Bayesian approaches give almost identical results for the estimated  coefficients $\hat{\beta_i}$, despite the addition of the prior. It seems that there is no preferred model, as indicated also by the comparison between  the corresponding AIC, BIC and the logarithm of  maximum likelihood ${\cal L}_\mathrm{max}$. 
We note though the smaller credible intervals from the Bayesian logit in comparison to those from the MLE analysis.

\subsection{The Pop III-Pop II/I dichotomy}\label{subsec:P2}

As previously mentioned, the first generation of stars (Pop III) are thought to form within pristine gas, while standard Pop II/I star formation takes place within metal  enriched gas. Here we investigate the Pop III-Pop II/I dichotomy using a Bayesian regression with logit and probit link functions.

\begin{figure}[ht]
\centering
\includegraphics[width=1\columnwidth]{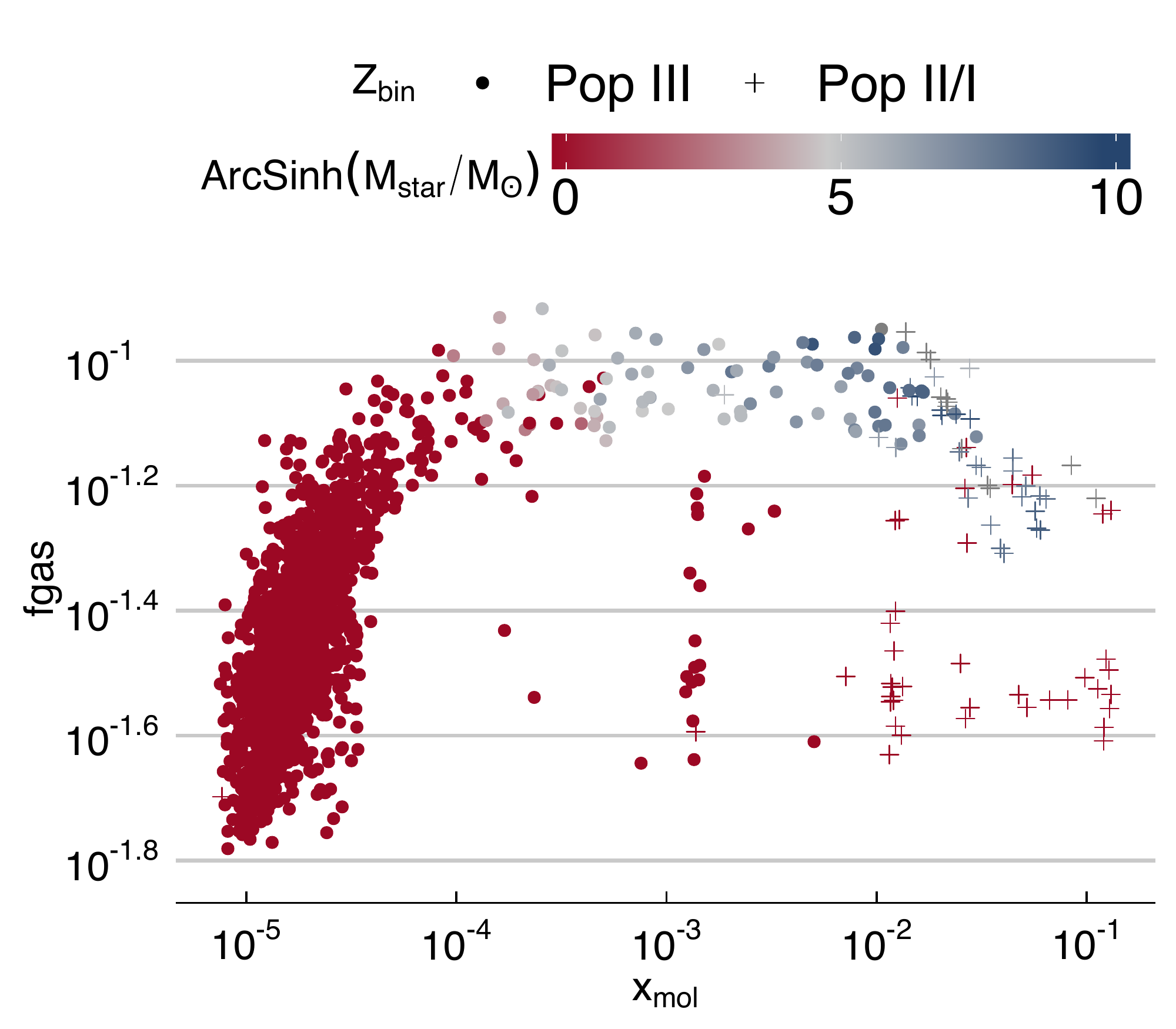}
\caption{
Scaled gas fraction versus molecular fraction, color-coded by stellar mass transformed by ArcSinh for visual purposes. Circles and crosses  represent pristine/low-metallicity Pop III and high-metallicity Pop II/I stellar environments, respectively.
}
\label{fig:Scat_P2}
\end{figure}

Figure~\ref{fig:Scat_P2} shows the gas fraction versus  molecular fraction with a colour scheme corresponding to stellar mass. A visual inspection indicates that larger molecular fractions are strongly associated with high-metallicity environments, confirming that the molecular fraction is the main predictor.
From a physical point of view, the fact that the gas fraction in the environment of Pop II/I stars is usually lower than that of Pop III stars suggests that the cosmological production of early heavy elements enhances significantly gas cooling capabilities and boosts molecule formation in polluted material well above $x_{\rm mol}\sim$ a few percents.
Basically, metal cooling allows gas fragmentation at regimes where pristine material is not able to condense -- see the region:
\{~$x_{\rm mol}>10^{-2}$, $f_{\rm gas} < 10^{-1.1}$~\}.

 The present cosmological simulations switch the stellar IMF  from top-heavy to standard Salpeter when the metallicity exceeds $Z_{\rm crit}=10^{-4}  Z_{\bigodot} $ (see \S \ref{sec:data}).
To perform the GLM analysis in this  section, we define $\Zbin$ as the binary response variable, depending on whether the gas metallicity lies above or below $Z_{\rm crit}$:
\begin{equation}
\Zbin = \left\{
 \begin{array}{rl}
  1 & \text{or `Pop II/I'~}\text{if } Z \geq Z_{\rm crit},\\
  0 & \text{or `Pop III'~} \text{if } Z < Z_{\rm crit}.
 \end{array} \right.
\end{equation}
One can then use the binomial GLM regression to determine which global galaxy properties are linked to the dichotomy between the Pop II/I and Pop III host environment and how. 
We also use this problem as an opportunity to demonstrate the use of both logit ($\eta=\log \pi/(1-\pi)$) and probit ($\eta=\Phi^{-1} (\pi)$) link functions.
Likewise the previous section, $\pi$ here represents the predicted probability for the success of the binary response variable, in other words if  a galaxy halo is an enriched Pop II/I environment given the underlying galaxy properties. 

Firstly, one must identify the key galaxy properties as predictor variables. As in the previous example, we scale the predictors by their respective means and divide by the standard deviations before performing the analysis. Nonetheless, rather than adding a set of pre-chosen predictors, we herein illustrate a general feature selection approach, making use of \texttt{step} function in \textsc{R}. The method attempts to alternately drop and add members of an input set of predictor candidates in order to minimize the AIC of the fitted model. By using the stepwise algorithm we are able to select the most parsimonious combination of parameters and interaction terms from our input set\footnote{See also the \texttt{drop1} function in \textsc{R}, which  is based on the likelihood ratio test.}.  The   scheme employed here falls into the category of wrapper methods of feature selection,  but there exist other approaches that can be  tailored to determine how relevant a feature is in representing a class in a high-dimensional space, i.e. so-called filter methods  \citep[see e.g., ][for a review of feature selection methods in astronomy]{Donalek2013}.

\begin{table}[ht]
\centering 
\caption{$\hat{\beta}_i$ coefficients from results of a Bayesian GLM analysis (with Cauchy prior) with logit and probit links. $\Zbin$, is the response variable, while the intercept $\hat{{\beta}}_0$, $\xmol$, \fgas, $M_{\mathrm{star}}$ and \stargas~ are predictors. The associated $p-values$ ($\cal P$) are listed underneath the coefficients. The logarithm of the maximum likelihood (${\cal L}_\mathrm{max}$), the AIC and the BIC for each choice of link function are also shown.} 
\label{tab:Coeff-P2}
\begin{tabular}{llcc} 
\\\hline 
\hline \\ [-2.2ex] 
 & & \multicolumn{2}{c}{\textit{Response variable:}} \\ 
\cline{3-4} 
\\[-2.2ex] & & \multicolumn{2}{c}{$\Zbin$} \\ 
\\[-2.2ex] & & Logit link	& Probit link \\ 
\hline \\[-2.5ex] 
\rowcolor{gray!10}
$\hat{{\beta}}_0$	
		&	& $-3.76 \pm 0.22$		& $-$1.94$ \pm 0.09$\\
\rowcolor{gray!10}           
  		&	& ($\cal P\ll$ 0.0001)		& ($\cal P\ll$ 0.0001)\\
       			& & \\ [-2.2ex] 
$\hat{\beta}_{1}$ &$\xmol$
			& 5.90$ \pm $0.65 	& 2.91 $\pm$ 0.32 \\ 
           	& & ($\cal P \ll$ 0.0001) 	& ($\cal P \ll$ 0.0001)\\ 
  			& & \\ [-2.2ex]
\rowcolor{gray!10}            
$\hat{\beta}_{2}$ & \fgas
			& $-0.97 \pm 0.27$	& $-0.43 \pm 0.12$ \\ 
\rowcolor{gray!10}             
  			& & ($\cal P$=0.0004) 		& ($\cal P$=0.0004)\\ 
 			& & \\ [-2.2ex] 
$\hat{\beta}_{3}$ &$M_{\rm star}$
			& $0.80 \pm 0.33$		& $0.54 \pm 0.22$ \\ 
  			& & ($\cal P$=0.02) 			& ($\cal P$=0.01)\\ 
  			& & \\ [-2.2ex]
\rowcolor{gray!10}            
$\hat{\beta}_{4}$ &\stargas
			& $-0.86 \pm 0.41$ 	& $-0.58 \pm 0.23$\\ 
\rowcolor{gray!10}             
  			& & ($\cal P$=0.04)	 		& ($\cal P$=0.01) \\ 
  			& & \\ [-2.2ex] 
\hline\\ [-2.2ex] 
${\cal L}_\mathrm{max}$	&	& $-$113 	& $-$114 \\ 
AIC	&				& 236 		& 238 \\ 
BIC & & 263 		& 265 \\ 
\hline 
\hline \\[-2.2ex] 
\end{tabular} 
\end{table} 

We found that $\xmol$ plays the most important role in the predictive power of the model. Furthermore, the factors that maximise the information gain and are worth including as predictor parameters are:
$\xmol$, \fgas,  \stargas,  and $M_{\mathrm{star}}$.
The selection is equivalent regardless of whether the logit or probit link functions are used.

Having chosen suitable input variables, we can then apply the GLM analysis as described in \S\ref{bayesLR} and \S\ref{bayesPR}. In Table~\ref{tab:Coeff-P2} we provide the estimated coefficients for the predictor variables and respective $p-values$. The coefficients in the two cases are different as can be seen in Table~\ref{tab:Coeff-P2}, which is mostly a consequence of the different choices of link function.  
Likewise, as in  section \ref{subsec:P1}, the logit coefficients  can be associated with probabilities. Once more,  $\xmol$ stands as the most influential variable,  and a variation  of $\approx 1.2\times 10^{-2}$ in $\xmol$,  increases the chances for an average  dark halo to be a potential host of Pop II/I stars by  a factor of 99.7\%.
The predicted probabilities $\pi$ are therefore  estimated by solving the following equation:
\begin{eqnarray}
\label{eqn:eta-P2}
\eta &=& 	\hat{\beta}_0 +\hat{\beta}_1 \xmol 
			+\hat{\beta}_2 f_\mathrm{gas} +\\
  &+&\hat{\beta}_3 M_{\mathrm{star}}+\hat{\beta}_4\frac{M_{\rm star}}{M_{\rm gas}}\nonumber, 
\end{eqnarray}
as well as either
\begin{eqnarray}
\label{eq:pi-P2-probit}
\Phi^{-1}(\pi) &=& \eta, 
\end{eqnarray}
if the probit link function (Eq. \ref{eq:probit}) is used, {\it or}
\begin{eqnarray}
\label{eq:pi-P2-logit}
\pi &=& \frac{e^{\eta}}{1+e^{\eta}}, 
\end{eqnarray}
for the logit link  function (Eq. \ref{eq:logit}).

\begin{figure}[h]
\centering
\includegraphics[width=1\columnwidth]{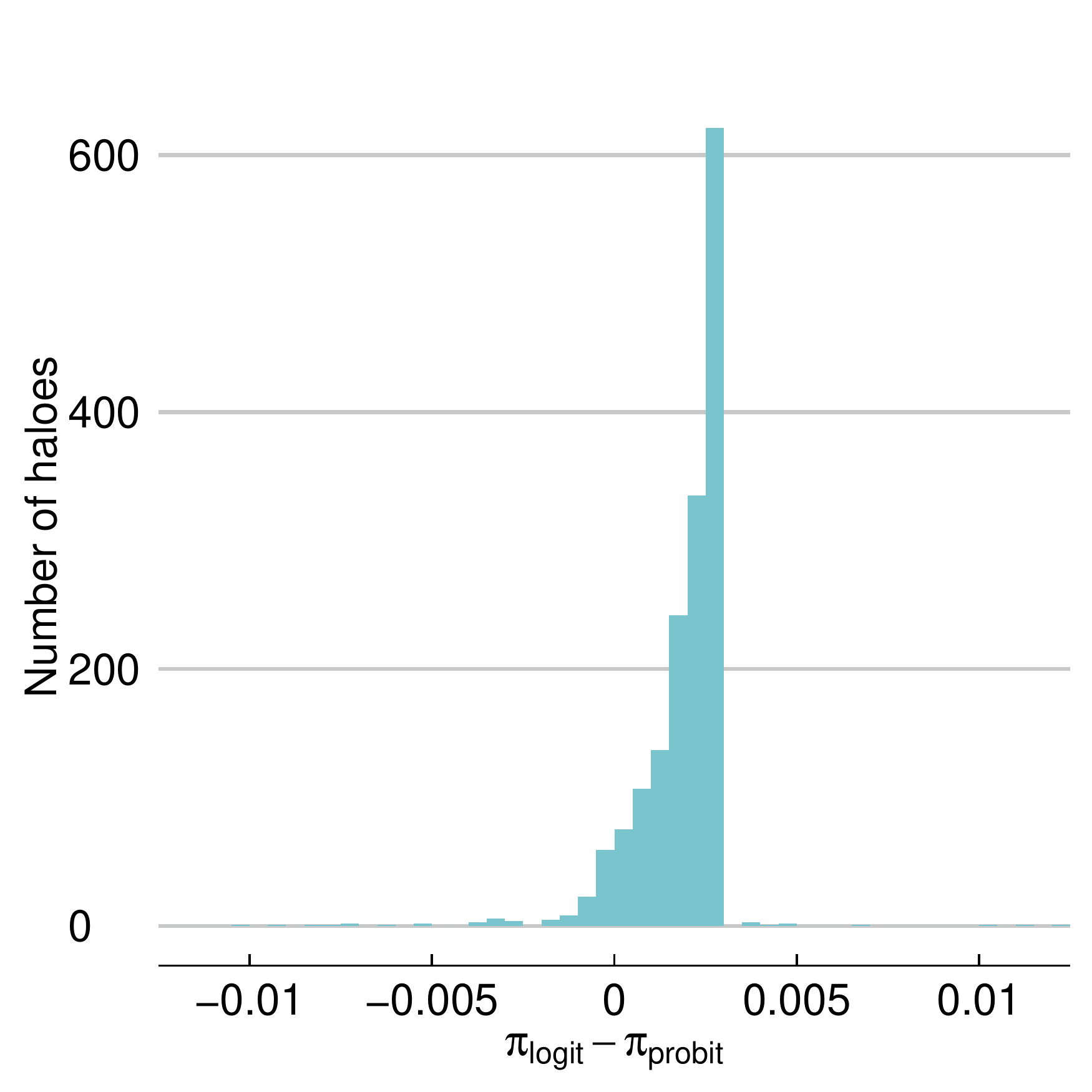}
\caption{
Histogram of the difference between predicted probabilities from logit, $\pi_{\rm logit}$,  and
probit $\pi_{\rm probit}$, regressions.}
\label{fig:pilogprob}
\end{figure}

Ultimately, logit and probit regression result in similar predictions for the probability that the response variable is unity, i.e. $\pi_\mathrm{logit}\approx\pi_\mathrm{probit}$. To illustrate this point, we calculate $\pi$ twice for each galaxy in our sample given its underlying properties: once using the logit link and then again using the probit link.
A histogram of the differences is shown in Figure~\ref{fig:pilogprob}.
The logit link function leads to a value of $\pi$ that is only  slightly higher. Thus,  for the case studied here, both link functions generate similar predictions, in spite of  their different interpretations \citep[see e.g.,][]{zuu13}.  
A  quantitative comparison between the predictive power of logit and probit, and the increase in number of {\it relevant} predictor variables are given in the following section.

\section{Diagnostics}\label{sec:diag}

We now describe our experimental setting to assess the performance of GLM on the prediction of star formation activity $\textit{SFR}_{\rm bin}$ and metal enrichment $Z_{\rm bin}$.
We report on accuracy (i.e., fraction of events correctly classified) using a resampling technique known as $10$-fold cross validation \citep{Hastie09} in \S \ref{subsec:cross}, on Receiver Operating Characteristic (ROC) curves \citep{Duda00} in \S \ref{subsec:ROC}  and on the confusion matrix in \S \ref{subsec:CM}.


\subsection{Cross validation} \label{subsec:cross}

When assessing model performance, it is of utmost importance to set aside a validation set to estimate the true generalization power of the model under analysis. This is particularly relevant to avoid the risk of model over-fitting. An over-fitted model captures aberrations on the training set that render the model useless during prediction. A popular approach to model validation makes use of resampling techniques \cite[][]{Hastie09}.

In the resampling technique known as $k$-fold cross validation, the data is divided into $k$ folds (subsamples) of equal size.
The technique runs iteratively as follows. On each  iteration, $k-1$ folds are used for training (model fitting), while the remaining fold is used for testing (model assessment). The procedure repeats $k$ times, using mutually exclusive testing folds across iterations. 
The final result is the average over the score obtained on each iteration.
Cross validation estimates the true performance of a classifier by exploiting all available information. 
In our experiments, we use a value of $k=10$ to achieve a trade-off between bias (proportional to $k$) and variance (inversely proportional to $k$).  Hereafter,  all  ROC curves and confusion matrices  are estimated  using the  $k=10$ cross-validation approach.


\subsection{ROC curves}\label{subsec:ROC}

ROC curves provide both a visually and quantitative approach to report on the accuracy of predictions for binary classifiers.
Hereafter,  we refer to the classifications as positive (1) or negative (0).
The technique consists of plotting the true positive rate (TPR or Sensitivity) vs the false positive rate (FPR or Specificity)
as we vary the decision boundary $\pi_{th}$. 
The variation in the decision boundary enables us to assess the performance of the classifier under unequal error costs (i.e., under scenarios where the cost of a false positive is different from a false negative).

Specifically, to generate a ROC curve we make use of two measurements: 
\begin{eqnarray}
\rm{Sensitivity} &=& \frac{\rm TP}{\rm TP + FN};\nonumber\\
\rm{Specificity} &=& \frac{\rm TN}{\rm TN + FP},
\end{eqnarray}
\noindent
where TP = true positives, FP = false positives, TN = true negatives, and FN = false negatives.
For example, in the case studied in \S \ref{subsec:P1} we would have:

\begin{itemize}
\item TP: the galaxy  has SF  and the method predicts  SF, 

\item FP: the galaxy  does not have SF  but the method predicts  SF, 

\item TN: the galaxy does not have SF  and the method predicts no SF,  

\item FN: the galaxy has SF,  but the method predicts no SF.  
\end{itemize}
In this case the Sensitivity (Specificity) would quantify the ability of the method to correctly identify galaxies with (without) SF: the closer to 1 these values are, the more successful the analysis is. 
The same interpretation holds for the case discussed in \S \ref{subsec:P2} by replacing $\textit{SFR}_{\rm bin}$ with $Z_{\rm bin}$. 

Sensitivity is normally plotted on the $y$-axis, while $1 -$ Specificity is plotted on the $x$-axis (Figs.~\ref{fig:ROC-Logit-ProbR} 
and~\ref{fig:ROC-GLM-NN}). 
The classifier is run several times with a different value of the decision threshold; each run provides a point in the (1-Specificity, Sensitivity) plane. The corresponding true ROC curve is obtained by joining the set of coordinates starting at (0,0) and ending at (1,1). An ideal ROC curve goes from (0,0) to (0,1) to (1,1). 
A quantitative approach to assess the quality of a ROC curve is to calculate the area under the curve (AUC), as a fraction of the area under the ideal curve, as often done in cases of discrepancy or inequality measurements \cite[since e.g.][]{gini1912, gini1921}.
Higher values of AUC correspond to more accurate classifiers, while a value of 0.5 corresponds to a random classifier \citep{Hil09}.

The ROC curve can be used to access the optimal  $\pi_{th}$, which is a trade-off between  Sensitivity and Specificity. In order words, it is the one corresponding to the coordinate with minimum distance from (0,1), where both Sensitivity and Specificity are maximum.  This is essential to ultimately assign a class membership for each data. A visual analysis of this classification scheme  is made via the confusion matrix, which will be discussed in the next section.

\subsection{Confusion Matrix}\label{subsec:CM}

A complementary diagnostics  method is the   confusion matrix $C$, which  captures information about the actual and predicted classifications of a particular learning algorithm or classifier \citep{kohavi1998}. 
 Columns   in $C$ correspond to actual classes, whereas rows  correspond to predicted  classes (e.g., $\textit{SFR}_{\rm bin}, Z_{\rm bin}$).
 The diagonal elements of the matrix contain the number of cases where the actual and predicted class agree, e.g., $C(i,i)$
contains the number of cases where class $i$ was predicted correctly. 
Off-diagonal elements capture all combination of misclassifications, e.g., $C(i,j)$ with $ i \ne j$
contains the number of cases where class $i$ was incorrectly predicted as class $j$.
On a $2 \times 2$ confusion matrix, entries along the diagonal stand for the number of true negatives TN (top left) and true positives TP (bottom right).
Specifically, $C$ can be represented as follows:

\begin{equation}
\begin{array}{c|c}
  \rm TN & \rm FN\\
  \hline
  \rm FP & \rm TP\\
 \end{array}.
\end{equation}

\section{Performance Comparison}\label{sec:comp}
During this section we compare the predictive performance of both logit vs probit links as discussed in \S \ref{subsec:P2} and between GLMs and artificial networks for the case discussed at \S \ref{subsec:P1}\footnote{Note that as the MLE and Bayesian approaches gives almost identical fitted coefficients, they lead to exactly same predicted probabilities.}.

\subsection{Logit vs Probit}
\begin{figure*}[t]
\centering
\includegraphics[width=1\columnwidth]{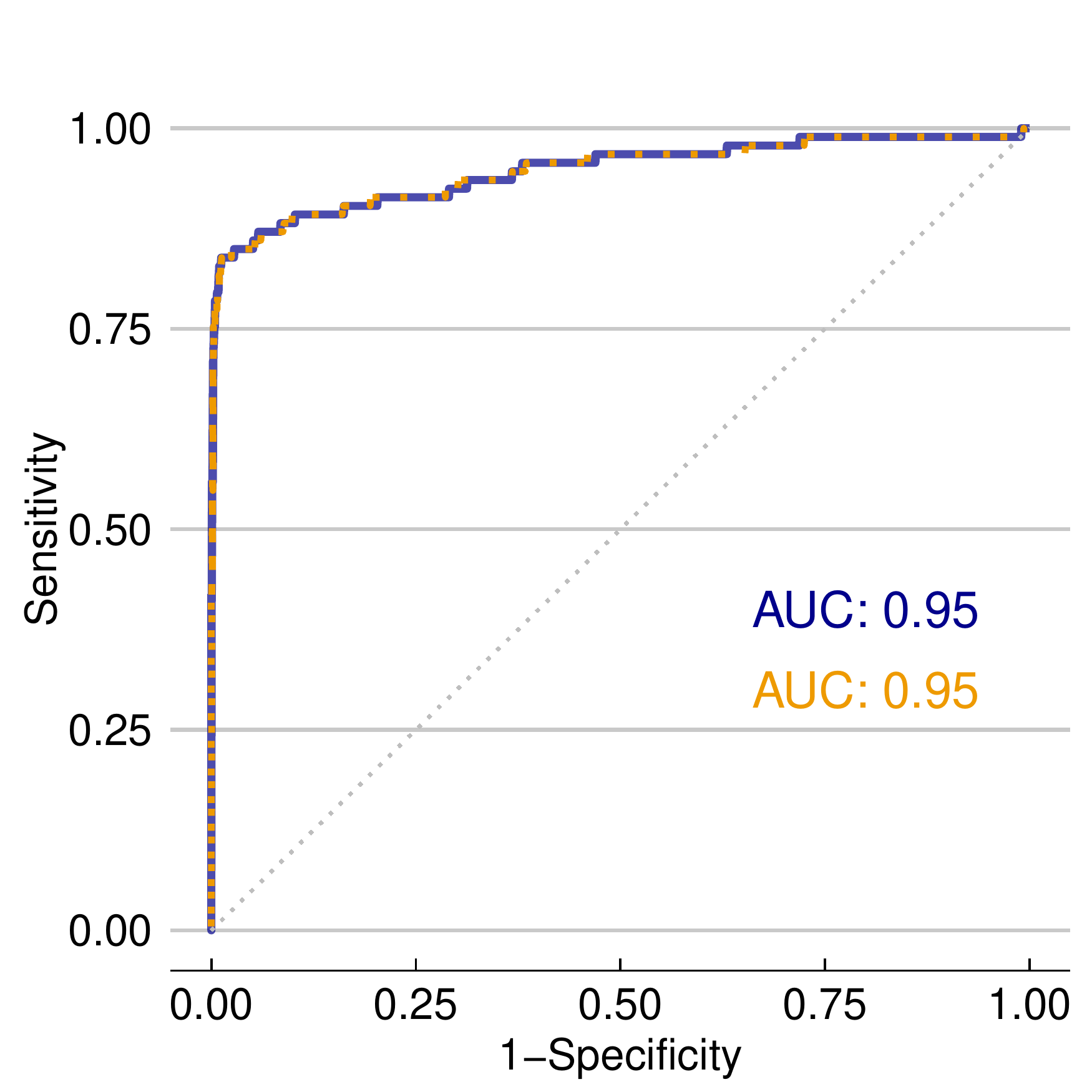}
\includegraphics[width=1\columnwidth]{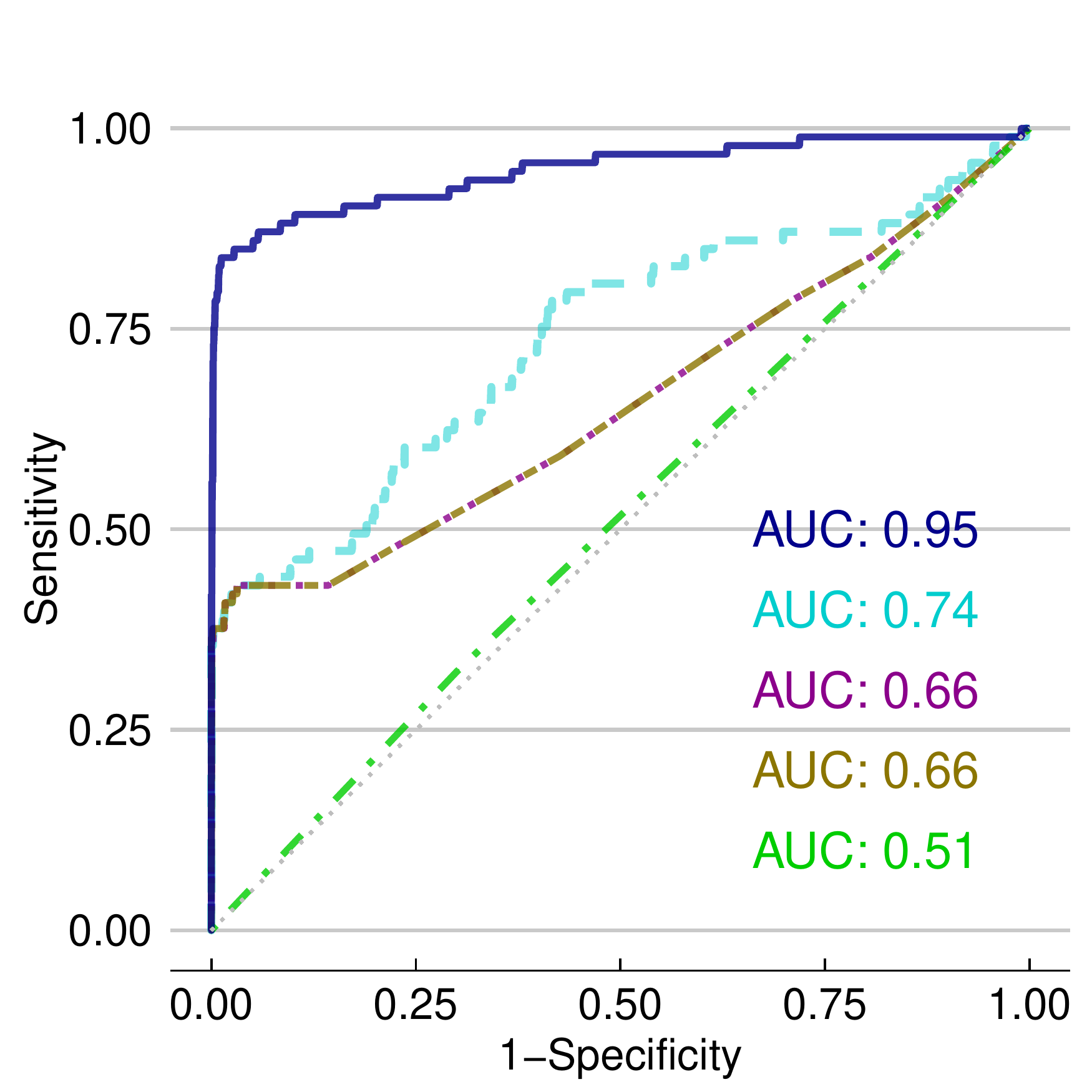}
\caption{
Left panel: ROC curves for logit (solid blue curve) vs probit (dotted orange curve) regression in the case discussed in \S \ref{subsec:P2}.
Right panel: ROC curves for logit regression for different combinations of predictor variables:
$\sim \hat{\beta_0}$ (dotted-dashed green curve);
$\sim \hat{\beta_0}+\hat{\beta_4}\frac{M_{\rm star}}{M_{\rm gas}}$ (dashed-dashed gold curve);
$\sim \hat{\beta_0}+\hat{\beta_4}\frac{M_{\rm star}}{M_{\rm gas}}+\hat{\beta_3}M_{\rm star}$ (dotted magenta curve); 
$\sim \hat{\beta_0}+\hat{\beta_4}\frac{M_{\rm star}}{M_{\rm gas}}+\hat{\beta_3}M_{\rm star}+\hat{\beta_2}f_{\rm gas}$ (dashed cyan curve);
$\sim \hat{\beta_0}+\hat{\beta_4}\frac{M_{\rm star}}{M_{\rm gas}}+\hat{\beta_3}M_{\rm star}+\hat{\beta_2}f_{\rm gas}+\hat{\beta_1}x_{\rm mol}$ (solid blue curve). The dotted grey  curve in both figures represents the performance of a random classifier. 
}
\label{fig:ROC-Logit-ProbR}
\end{figure*}

The left panel of Figure \ref{fig:ROC-Logit-ProbR} shows a comparison between logit and probit ROC curves, pointing to the equivalence in predictive power of both methods, achieving an outstanding performance of AUC = 0.95, although their coefficients have a different interpretation.

In  order to assess the relevance of a good set of predictor variables, the right panel of Figure \ref{fig:ROC-Logit-ProbR} shows a visualization of the logit GLM regression obtained adding different predictor variables. While $f_{\rm gas}$,  $M_{\rm star}$, and $M_{\rm star}/M_{\rm gas}$ together  have a non-negligible contribution to explain the metallicity enrichment  above/below $Z_{\rm crit}$ with an AUC = 0.74, the molecular fraction, $x_{\rm mol}$, clearly stands out as the most important parameter. This suggests that the level of  molecular gas fraction has a strong connection with the level of metal content in primordial haloes.

\subsection{Comparison between GLM and  Neural Networks}\label{sec:NN}

We compare GLM with a popular non-parametric technique to classification known as Artificial Neural Networks (ANN) \citep{Duda00}. A nonlinear multi-layer ANN is capable of expressing flexible decision boundaries over the variable space; it is a nonlinear statistical model that applies to both regression and classification. In particular, for an ANN with one hidden layer, each intermediate and output node computes a weighted combination of inputs, compressed (squashed) by a sigmoid (nonlinear) function \citep{Bishop96}.

\begin{figure}[h]
\centering
\includegraphics[width=1\columnwidth]{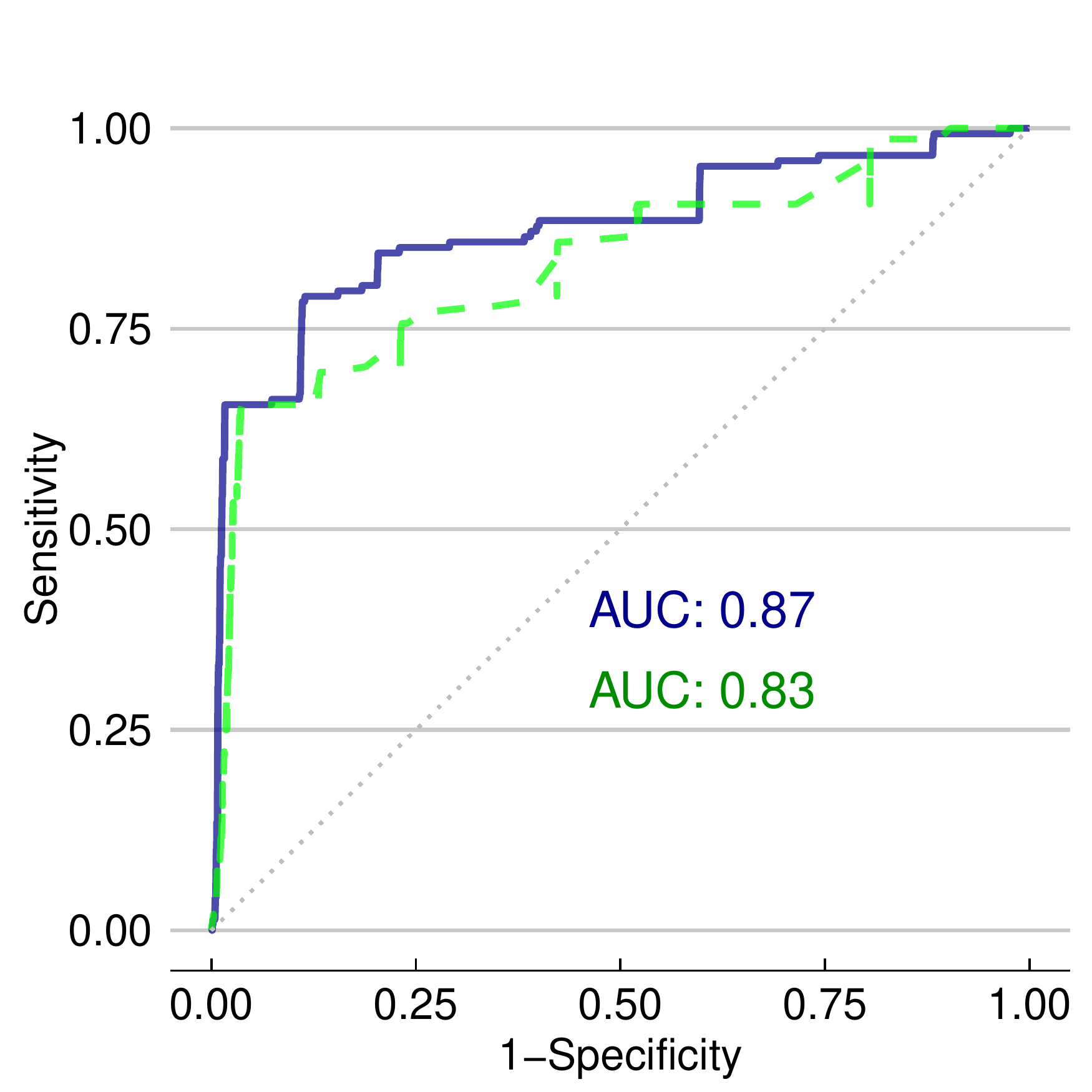}
\caption{ROC curves for Generalized Linear Model (solid blue curve) and Artificial  Neural Network (dashed green curve) for the case discussed in  \S \ref{subsec:P1}. The dotted grey   curve represents the performance of a random classifier.}
\label{fig:ROC-GLM-NN}
\end{figure}

Figure~\ref{fig:ROC-GLM-NN} shows ROC curves for GLM and ANN analysis of the case presented in \S \ref{subsec:P1}. 
The ROC curves were  generated
as those discussed in the previous section.
In our experiments, GLM attains an AUC slightly  higher than that of ANN (0.87 versus 0.83), reinforcing our claim for the competitiveness of GLM despite its inherent simplicity. We stress  that the above comparison should not be extrapolated to  imply  that GLMs are better suited for binary classification prediction  than ANNs, but for this particular and somewhat simple problem, both are equally good. The main advantages of GLMs are  their portability and  interpretation of the coefficients. Moreover, the  possibility to  approach the problem from a Bayesian perspective  is extremely  beneficial  when dealing with inherent issues of  observational data, such as errors-in-variables, selection bias, etc \citep{lor13}.

\begin{figure}[t]
{\includegraphics[width=0.475\columnwidth]{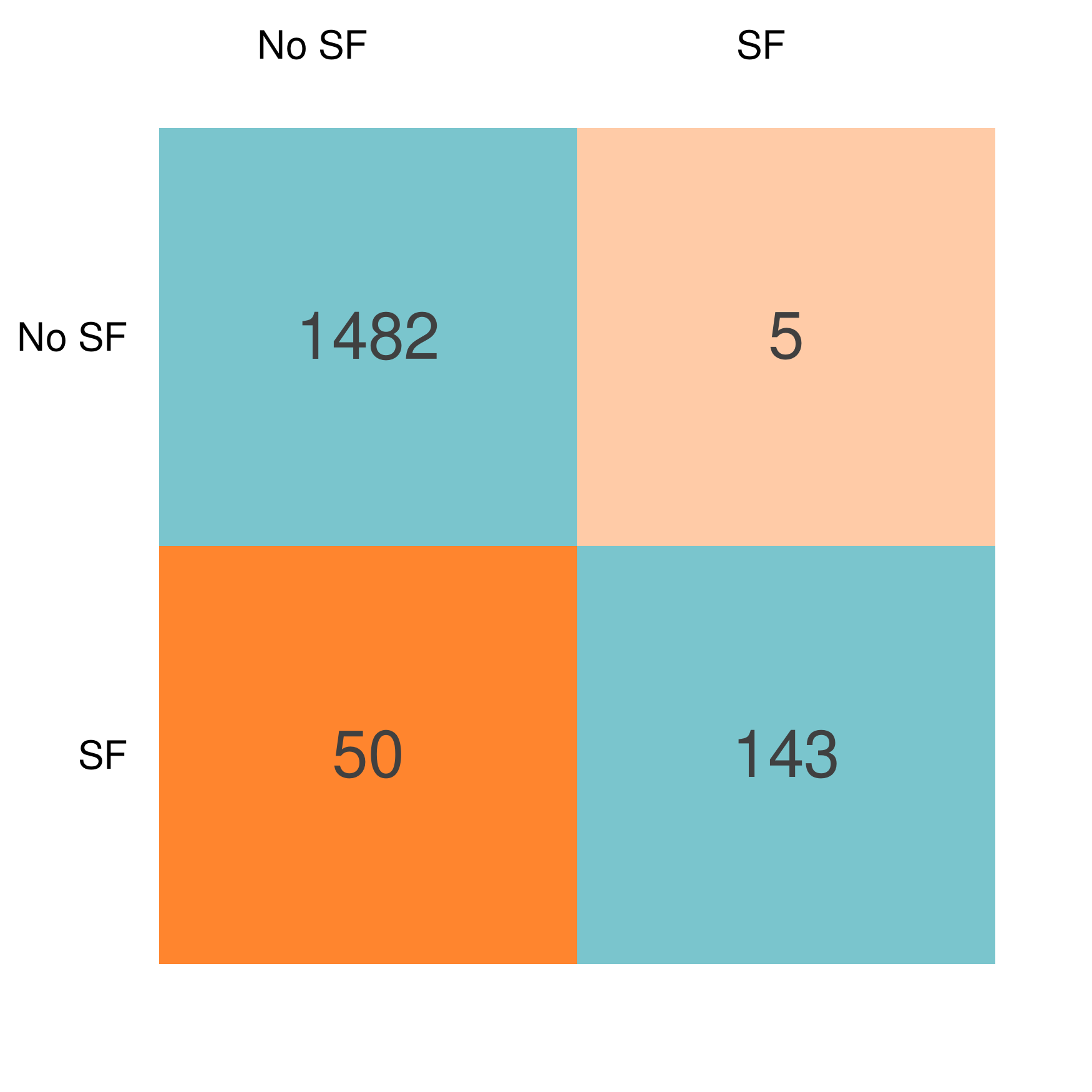}}
{\includegraphics[width=0.475\columnwidth]{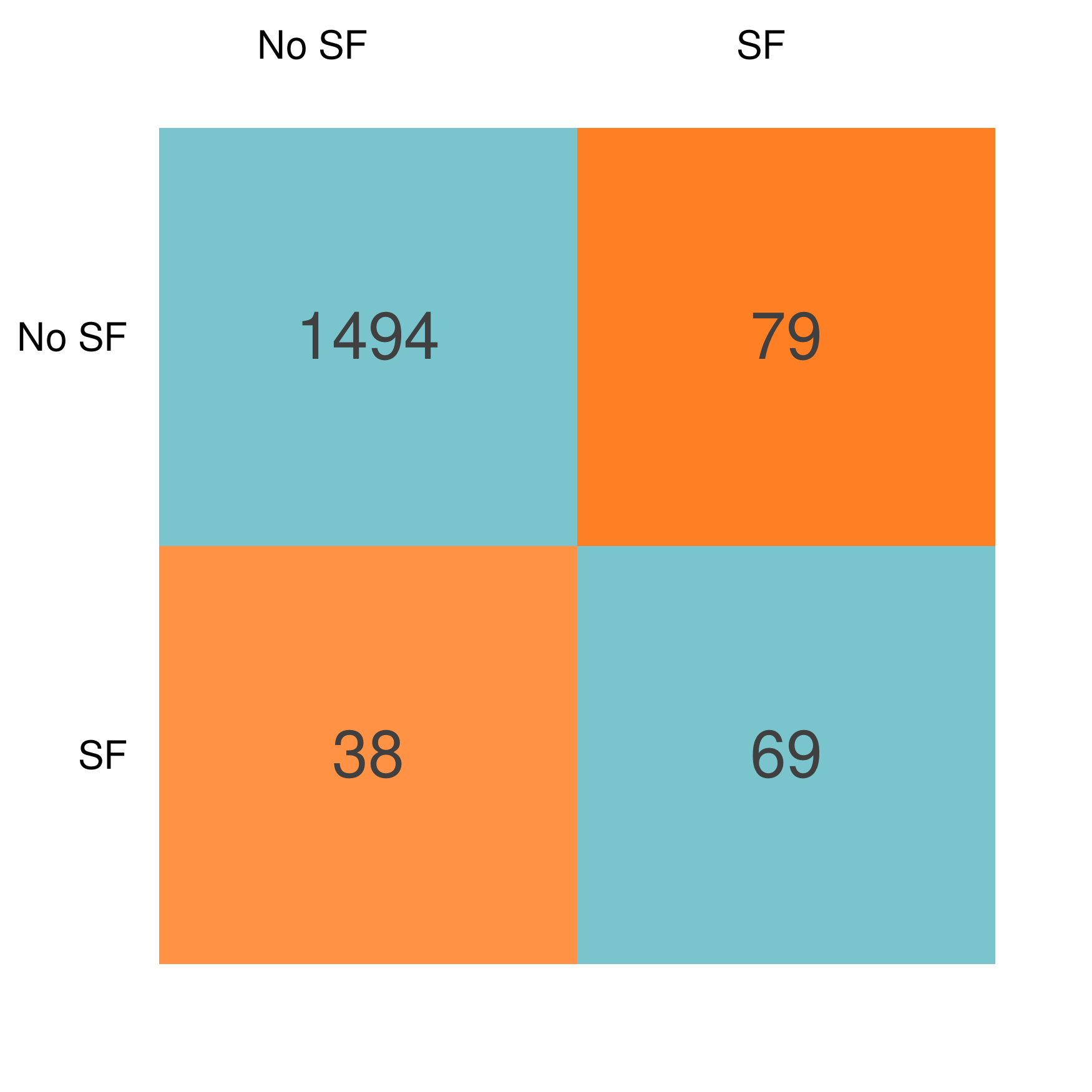}}
\caption{
Confusion matrix for the model SFR$_\mathrm{bin}\sim \xmol+Z$ (in {\sc R} notation) discussed in \S \ref{subsec:P1}, as expected by logistic GLM (left) and ANN (right).
In each panel, the first and second columns refer to the simulated objects with (1532 galaxies) and without (148 galaxies) star formation activity, respectively. The on-diagonal elements refer to TN (top left) and TP (bottom right), while the off-diagonal elements refer to FP (top right) and FN (bottom left).The color scheme ranges from blue, correct values,  to orange, incorrect values, with the intensity determined by the number of objects in each category. 
}
\label{fig:confusion-GLM-NN}
\end{figure}

Figure~\ref{fig:confusion-GLM-NN} shows two confusion matrices, one for GLM (left) and one for ANN (right) for the case underlined in \S \ref{subsec:P1}, i.e. the connection between star formation activity and the gaseous chemical properties $\xmol$ and $Z$ of galaxies.
While TN is similar under both classifiers, TP differs significantly:
ANN exhibits a high number of false negative FN (upper right) in contrast to the corresponding entry for GLM.
Hence, the overall accuracy\footnote{The accuracy is given by $(TN+TP)/(TN+TP+FP+FN).$} of GLM is $96.7\%$, while that of ANN is $93\%$.

\section{Conclusions}
\label{sec:end}
%
We perform a comprehensive introduction of logit and probit  generalized linear model regression for the astronomical community from both a maximum likelihood and a Bayesian perspective. 
As a real application, we analyse the host environment of the first generation of stars as predicted by numerical hydro-simulations of the early Universe, including detailed chemistry, gas physics, star formation, stellar evolution and stellar feedback.
A summarizing flowchart visualization of the entire process is given in \ref{sec:appendix}.

The halo  properties analysed here are categorical with two possible outcomes and therefore ideal candidates for the application of binomial GLM regression.
These correspond to either (i) the presence/absence of star formation activity, or (ii) metal content above/below the critical metallicity associated to stellar population transition in primordial epochs.

In the first case, the explanatory variables were decided beforehand with preliminary physical motivation, while in the second case, we demonstrated the use of the AIC to select the most parsimonious set of variables from among a given set of candidates. This method is particularly beneficial for providing new insight into fundamental underlying galaxy properties.

A maximum likelihood as well as a Bayesian (with Cauchy priors) analysis result in very similar coefficients for each variable.  
We have explored the use of both logit and probit link functions and found that they lead to different $\hat\beta$ coefficients, but with the same sign.
Nevertheless, calculations of the predicted probabilities produce very similar results regardless of whether a logit or probit model is used for estimation.

The GLM method has been shown to be very competitive against artificial neural networks, attaining an area under the curve (AUC) coefficient of 0.87 against 0.83 from ANN.
Since a value of $\rm AUC = 1$ indicates a perfect classifier and a value of $\rm AUC = 0.5$ suggests a random predictor, both GLM and ANN approaches can be considered rather robust, albeit the AUC seems to favour slightly the GLM for this particular test.
Furthermore, given its inherently simplicity, GLM results are easily portable and have a more straightforward interpretation of its coefficients in terms of odds and probabilities.

Also worth noting is that the potential of GLM regression goes far beyond binary classification.
Many data situations involve discrete data, but are nevertheless modelled as if the response variable were continuous.
If the data are modelled as discrete, it is by employing 
a Poisson model, without due regard for the corresponding distributional assumption of equality between mean and variance (equidispersion).
This is a strongly restrictive technical assumption and is rarely met in real data.
In practice, there are nearly as many count models as there are shapes of counts: there is a variety of mixture models, of zero-inflation models, of two-part hurdle models, of finite mixture models, etc. which assume that the counts being modelled are being generated from more than one source.
There are situation where the  classic GLM assumption of uncorrelated measurements fails, e.g., for repeated measurements from the same object. For theses cases,  plenty of  extensions exist,  such as generalized estimating equations \citep{lia86}. Additionally, there are generalized additive models,  non-parametric quantile count models,  models with endogenous stratification, panel models, and   3-parameter count models, to name only a few.
GLMs are of common use in the statistical literature, but 
almost {\it Terra incognita} in astronomical data analysis, with only few recent notable applications of logistic regression (e.g.\ \citealt{Rai12,Rai14,lan14}), Poisson regression (e.g.\ \citealt{and10}) and  negative binomial regression \citep{ata14}.

Finally, we highlight the vast potential of GLMs and extended GLMs for the astronomical community through their possible application to a plethora of astronomical problems, such as: photometric redshift estimation \citep[gamma distributed data;][]{Elliott2015}, globular cluster counts (Poisson distributed data), or galaxy morphological classification (multinomial distributed data).
GLMs might be a precious instrument for astronomical investigations, thanks to their capabilities in addressing scientific questions that could not be answered otherwise.
Thus, we are confident in a prompt integration of these methods into astronomy, with the hope that contemporary statistical techniques may become common practice in the $\rm 21^{st}$ century astrophysical research.

\section*{Acknowledgements}
We thank the referee for very useful comments that helped to improve this manuscript. 
We thank M. L. L. Dantas for   the careful review and fruitful comments of  the manuscript. MK acknowledges support by the DFG project DO 1310/4-1.
UM would like to thank funding from a Marie Curie Fellowship of the European Union Seventh Framework Project  (FP7/2007-2013), grant agreement n. 267251.
Work on this paper has substantially benefited from using the collaborative website AWOB (http://awob.mpg.de) developed and maintained by the Max-Planck Institute for Astrophysics and the Max-Planck Digital Library.
The bibliographic research was possible thanks to the tools offered by the NASA Astrophysical Data Systems and the JSTOR archive.

\appendix
\section{\textsc{R} scripts}
\label{sec:scripts}

In the following, we display the \textsc{R} scripts for the models discussed in sections \ref{maxlikeGLM},\ref{bayesLR}, and \ref{bayesPR}, respectively. 

\subsection*{MLE with logit link }
The basic syntax for a MLE logit model:


\begin{lstlisting}
    glm.fit <- glm(y~x1+x2+...,
    family = binomial("logit")).
\end{lstlisting}
\
The \texttt{summary} command can be called on the \texttt{glm.fit} object returned, as can \texttt{plot} which will display a number of useful fit and model checking diagnostics. 

\subsection*{Bayesian GLM with  logit link }
The basic syntax for a Bayesian logit model:
\begin{lstlisting}                
   library(arm) 
  #Output identical to ML logit
   blr1 <- bayesglm(y ~ x1+x2+ ..., 
   family=binomial(link="logit"),
   prior.scale=Inf, prior.df=Inf,
   data=<datafile>)
   display(blr1)          
                
  #Bayes GLM  with default binomial
  #logit link and  Cauchy prior
  #with scale=2.5
   
   blr2 <- bayesglm(y~x1+x2+...,          
   family=binomial,
   data=<datafile>)
   display(blr2)
   
  #Bayes logit with normal prior 
  #with scale=2.5                    
   blr3 <- bayesglm(y~x1+x2+..., 
   family=binomial,
   prior.scale=2.5, prior.df=Inf,
   data=<datafile>)
   display(blr3). 
\end{lstlisting} 
\

\subsection*{Bayesian GLM with  probit link }
The basic syntax for a Bayesian probit model:

                
 \begin{lstlisting}   
   library(arm)
    bpr <- bayesglm(y~x1+x2+...,
    family=binomial(link="probit"), 
    prior.scale=2.5, prior.df=Inf,
    data=<datafile>)
    display(bpr).
\end{lstlisting}    
\

\section{Flowchart  for  GLM regression}
\label{sec:appendix}
     \tikzstyle{io} = [trapezium, trapezium left angle=70, trapezium right angle=110, text width=6em, text centered,draw=black,                       fill=green!35]
     \tikzstyle{dia} = [diamond, draw, fill=red!40, 
                            text width=4.5em, text badly centered, node distance=3cm, inner sep=0pt]
     \tikzstyle{block} = [rectangle, draw, fill=blue!40, 
                         text width=5em, text centered, rounded corners, minimum height=4em]
     \tikzstyle{line}  = [draw, -latex']
     \tikzstyle{line2}  = [draw, -latex',out=-20,in=-70]
     \tikzstyle{cloud} = [draw, ellipse,fill=gray!25, node distance=3cm,
                         minimum height=2em]
     \tikzstyle{polygon}= [draw,regular polygon,regular polygon sides=7,fill=orange!35,text  centered,text width=3.75em,
                            minimum height=1em]

This section illustrates a brief summary of GLM analysis and model  diagnostics. It comprises:
\begin{itemize}
\item Acquire the dataset.
\item Choose the response variable  to be modelled. 
\item Choose predictor variables.
\item Choose  GLM family, e.g. Gaussian, Poisson, binomial.
\item Choose  either a maximum-likelihood or a  Bayesian approach.
\item Choose  link function.
\item Estimating coefficients by means of a GLM or Bayesian GLM analysis, i.e., estimate $\eta$ and predicted probabilities $\pi$
\item Classification and diagnostic tests:
\begin{itemize}
\item ROC curve-probability threshold.

\item Confusion Matrix for a given $\pi_{th}$ and assigned class memberships.
\end{itemize} 
\end{itemize}
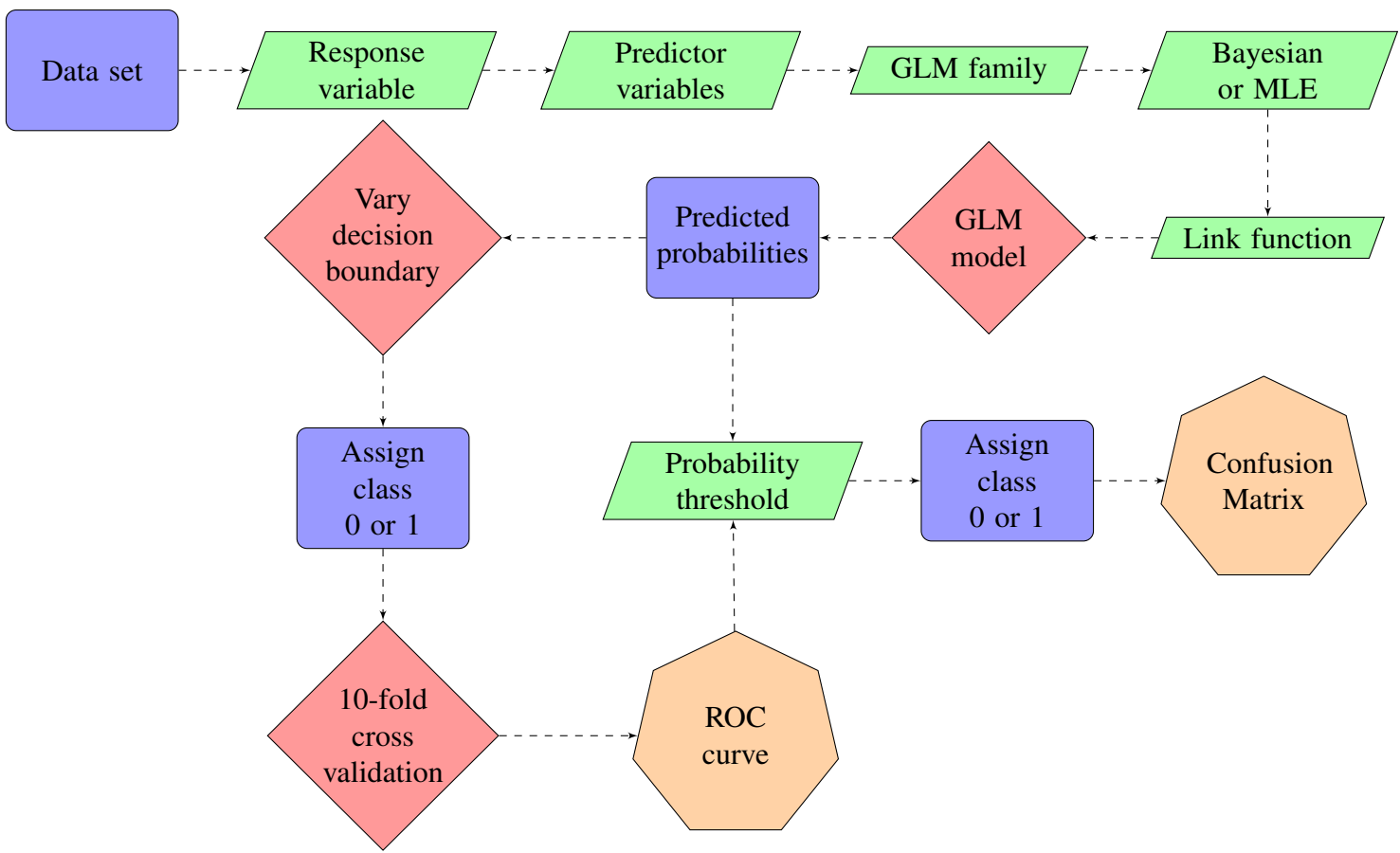
\begin{figure*}
\begin{tikzpicture}[node distance = 2cm, auto]
   
    \node [block] (train) {Data set};
       \node [io, right = 1cm  of train] (select0) {Response  variable};
    \node [io, right = 1cm  of select0] (select1) {Predictor variables};
    \node [io, right = 1cm  of select1] (glm) {GLM family};
    \node [io, right = 1cm  of glm] (method) {Bayesian or MLE};
    \node [io, below = 1.5cm  of method] (link) {Link function};
    \node [dia, left  = 1 cm  of link] (fit) {GLM model};
        \node [block, left = 1cm  of fit] (prob) {Predicted probabilities}; 
        \node [io, below  = 2 cm  of prob] (tau) {Probability threshold };
    \node [block, right = 1cm  of tau] (class) {Assign class\\ 0 or 1}; 
    \node [polygon, right = 1 cm  of class] (confusion) {Confusion Matrix}; 
       \node [dia, left  = 2 cm  of prob] (var) {Vary decision boundary };
        \node [block, below = 1cm  of var] (class2) {Assign class\\ 0 or 1}; 
        \node [dia, below  = 1 cm  of class2] (cross) {10-fold cross validation};
    \node [polygon, right = 1.925cm  of cross] (roc) {ROC curve};  
  
    
    \path [line,dashed] (train)--(select0);
     \path [line,dashed] (select0)--(select1);
    \path [line,dashed] (select1)--(glm);
    \path [line,dashed] (glm)--(method);
    \path [line2,dashed] (method)--(link);
    \path [line,dashed] (link)--(fit);
    \path [line,dashed] (fit)--(prob);
       \path [line,dashed] (prob)--(tau);
        \path [line,dashed] (tau)--(class);
       \path [line,dashed] (class)--(confusion);
    \path [line,dashed] (prob)--(var);
    \path [line,dashed] (var)--(class2);
    \path [line,dashed] (class2)--(cross);
     \path [line,dashed] (cross)--(roc);
        \path [line,dashed] (roc)--(tau);
\end{tikzpicture}
\caption{Tabular data is represented by blue rectangles, calculations by red diamonds, choices  by green parallelograms, and diagnostic outcomes by orange heptagons.}
\end{figure*}


\end{document}